\documentclass[]{aa}
\usepackage{graphicx}
\usepackage{subfigure}
\usepackage{txfonts}
\usepackage{natbib}
\bibpunct{(}{)}{;}{a}{}{,} 

\pdfoutput=1

\newcommand{\solarmass}{\mathrm{M}_{\sun}}

\begin{document}

\title{Galaxies undergoing ram-pressure stripping: the influence of the bulge on morphology and star formation rate}

\author{{D. Steinhauser \inst{1}}
          \and
           {M. Haider \inst{1}}
          \and
	   {W. Kapferer \inst{1}}
	  \and 
           {S. Schindler \inst{1}}}

   \offprints{Dominik Steinhauser, \email{dominik.steinhauser@uibk.ac.at}}

   \institute{\inst{1}Institute of Astro- and Particle Physics, University
   of Innsbruck, Technikerstr. 25, 6020 Innsbruck, Austria
   }

   \date{-/-}

   \titlerunning{Simulations of galaxies in a cluster environment}

  \abstract
   {}
   {We investigate the influence of stellar bulges on the star formation and morphology of disc galaxies that suffer from ram pressure. Several tree-SPH (smoothed particle hydrodynamics) simulations have been carried out to study the dependence of the star formation rate on the mass and size of a stellar bulge. In addition, different strengths of ram pressure and different alignments of the disc with respect to the intra-cluster medium (ICM) are applied.}
   {The simulations were carried out with the combined N-body/hydrodynamic code GADGET-2 with radiative cooling and a recipe for star formation. The same galaxy with different bulge sizes was used to accomplish 31 simulations with varying inclination angles and surrounding gas densities of $10^{-27}\, \mathrm{g}\,\mathrm{cm}^{-3}$ and $10^{-28}\,\mathrm{g}\,\mathrm{cm}^{-3}$. For all the simulations a relative velocity of $1000\,\mathrm{km}\,\mathrm{s}^{-1}$ for the galaxies and an initial gas temperature for the ICM of $10^7\,\mathrm{K}$ were applied. Besides galaxies flying edge-on and face-on through the surrounding gas, various disc tilt angles in between were used. To allow a comparison, the galaxies with the different bulges were also evolved in isolation to contrast the star formation rates. Furthermore, the influence of different disc gas mass fractions has been investigated.}
   {As claimed in previous works, when ram pressure is acting on a galaxy, the star formation rate (SFR) is enhanced and rises up to four times with increasing ICM density compared to galaxies that evolve in isolation. However, a bulge suppresses the SFR when the same ram pressure is applied. Consequently, fewer new stars are formed because the SFR can be lowered by up to $2\,\solarmass\, \mathrm{yr}^{-1}$. Furthermore, the denser the surrounding gas, the more inter-stellar medium (ISM) is stripped. While at an ICM density of $10^{-28}\,\mathrm{g}\,\mathrm{cm}^{-3}$ about 30\% of the ISM is stripped, the galaxy is almost completely (more than 90\%) stripped when an ICM density of $10^{-27}\mathrm{g}\,\mathrm{cm}^{-3}$ is applied. But again, a bulge prevents the stripping of the ISM and reduces the amount being stripped by up to 10\%. Thereby, fewer stars are formed in the wake if the galaxy contains a bulge. The dependence of the SFR on the disc tilt angle is not very pronounced. Hereby a slight trend of decreasing star formation with increasing inclination angle can be determined. Furthermore, with increasing disc tilt angles, less gas is stripped and therefore fewer stars are formed in the wake. Reducing the disc gas mass fraction results in a lower SFR when the galaxies evolve in vacuum. On the other hand, the enhancement of the SFR in case of acting ram pressure is less pronounced with increasing gas mass fraction. Moreover, the fractional amount of stripped gas does not depend on the gas mass fraction.}
   {}


   \maketitle
%

\section{Introduction}

As is known since the 1970s, the star formation rate of galaxies depends heavily
on their environment. As stated by \citet{bo78}, cluster galaxies are found to be redder than field galaxies.  
This implies that cluster galaxies must form stars at
a lower rate than their counterparts in the field. Although the star formation rate is lower in 
cluster galaxies, they also tend to have less gas than galaxies in less dense 
environments. \\
Furthermore, an excess of blue objects has been found at redshift
$z = 0.5$, compared to lower redshifts. Moreover, spiral galaxies have a higher star formation rate than ellipticals. 
Ellipticals on the other hand are more numerous in the centre of a galaxy cluster \citep{bark09}. Furthermore,
\citet{dressler97} found that between redshift $z = 0.5$ and $z = 0$ the fraction of S0 galaxies is strongly 
increasing. Hence, much more early-type galaxies can be found in clusters, which could be explained by
different galaxy transformation processes in clusters of galaxies. \\
\noindent
Many different mechanisms have been studied in this context to explain the change in morphology and 
star formation of cluster galaxies. Among others, galaxy harassment \citep[e.g.][]{moore98} and 
strangulation/starvation \citep[e.g.][]{larson80} are discussed mechanisms. Furthermore, galaxy-galaxy 
interactions \citep[e.g.][]{sulentic76, bushouse87, sosa88, kewley06}
as well as ram-pressure stripping \citep{gunn_gott72} can be found. Because these different processes
can happen at the same time, it is crucial to separate these processes which can be done by
studying the evolution of the star formation rate in dependence on the environment \citep[e.g.][]{poggianti06}. \\
\noindent
Clusters of galaxies contain a hot ($\approx 10^7\,\mathrm{K}$) and very thin ($10^{-29} - 10^{-26} \, \mathrm{g}\,\mathrm{cm}^{-3}$) gas, 
the so-called intra-cluster medium, detected by X-ray observations. Galaxies that are moving in the cluster's potential and 
consequently are exposed to the ICM, are feeling the ram pressure of this hot gas and the inter-stellar medium 
is stripped if the ram pressure exceeds the gravitational restoring force, according to the Gunn and Gott criterion \citep{gunn_gott72}.
Additionally, ram-pressure stripping is also a mechanism for enriching the ICM with metals \citep[e.g.][]{schindler08, lovisari09, durret11}. The influence of ram-pressure stripping on the morphology and star formation was exhaustively investigated in the last decade
\citep[e.g.][]{gavazzi01, yoshida04, scott10, kenney04}, by investigating HI deficiencies and $\mathrm{H}\alpha$ and CO observations to investigate regions of star formation. This showed that interactions with the ICM cause the galaxies to have morphological disturbances
of the gas disc and extensive gas loss in the central part of a galaxy cluster \citep[e.g.][]{boselli09}. In addition, star formation has also been found outside the galaxies 
themselves \citep[e.g.][]{hester10}. \\
\noindent
Another approach to study the different interaction processes of galaxies in a cluster are numerical simulations. Tidal interactions between galaxies 
and galaxy-galaxy mergers were investigated by \citet{kronberger06, kronberger07}. Additionally, they provided studies of the influence on the rotation 
curves and velocity fields of spiral galaxies \citep{kronberger07, kapferer06}. Furthermore, it has been found that nonaxisymmetric and 
nonbisymmetric features are introduced by tidal interactions \citep{rubin99}. Ram-pressure stripping simulations were also performed by several 
groups, which found a good agreement of the Gunn \& Gott criterion in hydrodynamic simulations as well \citep{abadi99, vollmer01, roediger05}. \citet{jachym09, jachym07} investigated the influence of a time-varying ram pressure and different inclination angles of
the galaxies relative to the movement direction of a galaxy. These simulations were performed mostly with a smoothed particle hydrodynamics (SPH) prescription. Other methods, including sticky particles \citep[e.g.][]{vollmer01} and Eulerian grid-codes \citep[e.g.][]{roediger06, roediger07} were also used to simulate ram-pressure stripping. \\
\noindent
Because one of the main aims of this paper is to study the evolution of the star formation rate depending on the environment, the simulation
code includes a star formation recipe.
A high pressure exerted by the ICM to the ISM can enhance the star formation rate significantly, as e.g. 
found by \citet{kronberger08}.
Other simulations also confirm the enhancement of the star formation rate when ram pressure is applied. \citet{bekki03}
demonstrated that the pressure of the ICM can induce the collapse of molecular clouds and consequently trigger a burst of star 
formation. Also using SPH, their recipe for star formation employs the local dynamical timescale and the sound crossing time of
the gas particles to transform them into stellar particles. Furthermore, \citet{bekki11} used a Schmidt-law with exponent $\gamma = 1.5$
to control the rate at which new stars are produced. \citet{vollmer12} also use a model for star formation in their simulation 
prescription, adopting the sticky-particles approach for the hydrodynamics with ram pressure applied by an additional acceleration
to the particles. Stars are being formed during cloud collisions, represented by tracer-particles with zero mass. 
These particles are then evolved passively. \\
In our approach, star formation processes are not restricted to the galaxy or some parts of the galaxy
itself but are separately calculated for each particle, considering temperature and density in a self-consistent way throughout the whole simulation domain.
In the same way, new stellar particles are spawned that conserve the phase-space properties of the gas particle, containing the mass of 
newly formed stars. Thus, enhancements in the star formation rate due to ram pressure are obtained self-consistently.
Already adopted for ram-pressure stripping simulations in \citet{kronberger08}, we describe this recipe in more detail
in section \ref{sim_setup}. \\
\noindent
In this work, especially the influence of the combination of a galaxy's morphology and ram-pressure stripping 
on the star formation rate is investigated. In this context, the special case of the influence of a stellar bulge is studied. We use different 
strengths of ram pressure and different model galaxies to decode a bulge's influence on ram-pressure stripping. The paper is organised as follows: in Sect. 2 we present the numerical setup. In Sect. 3 the results are presented and are compared to observations in Sect. 4. Finally, in Sect. 5 a summary with the main conclusions is given. \\

\section{Numerical setup}

\subsection{Simulation implementation}
\label{sim_setup}

All simulations in this paper were performed with the N-body/hydrodynamic code GADGET-2 developed by Volker Springel \citep{springel05}. The N-body part treats the collisionless dynamics of the dark matter and the stellar component. The gravitational force computation is performed using direct summation for short range forces and a sophisticated tree code \citep[as first introduced by][]{bh86} and treePM code for the long range forces. The gas of the galaxies and the ICM is treated hydrodynamically via SPH \citep[][]{gingold77, lucy77}. \\
Furthermore, the version of GADGET-2 used here includes radiative cooling \citep{katz96}, a recipe for star formation, stellar feedback and galactic winds as described in \citet{springel03}. This so-called hybrid model for star formation was already applied in \citet{kronberger08} and \citet{kapf09, kapf10}. Because the star formation is of particular interest in this paper, we present the main properties of the star formation model in the following. \\
The star formation in the ISM is spatially not resolved but is described in a statistical manner. The continuous matter distribution is represented by SPH particles with spatially averaged hydrodynamic quantities. The star formation is thus calculated for each such particle. It is assumed that stars form from cold gas clouds, which cannot be resolved in the simulations either. A certain fraction of mass of a single SPH particle is then considered as cold gas clouds and the total gas density can be partitioned as $\rho = \rho_c + \rho_h$, namely the cold gas clouds and the hot ambient medium. From these cold gas clouds, the stars are formed on a characteristic timescale $t_\star$, whereas a certain mass fraction $\beta$ is again immediately released in the form of supernovae from massive stars with $M > 8\;\solarmass$. As in the other works mentioned before, we adopt $\beta = 0.1$ according to a Salpeter IMF with a slope of $-1.35$ with lower and upper limit of $0.1\,\solarmass$ and $40\,\solarmass$ respectively. Then, the simple model of star formation and hence the time evolution of the stellar density $\rho_\star$ is

\begin{equation}
	\frac{\mathrm{d}\rho_\star}{\mathrm{d}t} = \frac{\rho_c}{t_\star}-\beta\frac{\rho_c}{t_\star} = \left( 1 - \beta\right) \frac{\rho_c}{t_\star}\;.
\end{equation}

\noindent In addition to mass release, energy is also injected into the ISM by supernovae. The ambient gas is heated by this feedback energy and the cold gas clouds can be evaporated. On the other hand, gas cools through the implemented radiative cooling process. A minimum temperature of $10^4 \mathrm{K}$ can be reached. Therefore, a redistribution of the two different gas phases takes place. The density distribution of hot and cold gas evolving in time can consequently be written as

\begin{equation}
	\frac{\mathrm{d}\rho_c}{\mathrm{d}t} = - \frac{\rho_c}{t_\star} - A\beta\frac{\rho_c}{t_\star} + \frac{1-f}{u_h - u_c}\Lambda_{\mathrm{net}}\left( \rho_h, u_h\right)
\end{equation}

\noindent and 

\begin{equation}
 	\frac{\mathrm{d}\rho_h}{\mathrm{d}t} = \beta\frac{\rho_c}{t_\star} + A\beta\frac{\rho_c}{t_\star} - \frac{1-f}{u_h - u_c}\Lambda_{\mathrm{net}}\left( \rho_h, u_h\right)
\end{equation}

\noindent for hot ($\rho_\mathrm{h}$) and cold ($\rho_\mathrm{c}$) gas density, respectively, with $u$ being the internal energy, $A$ the efficiency of the evaporation process (depending on the density) and $\Lambda_{\mathrm{net}}$ the cooling function \citep{katz96}. $f$ is used to distinguish between ordinary cooling ($f = 1$) and thermal instability ($f = 0$) when star formation is going on. The switch to thermal instability cooling is made using a simple density threshold $\rho_\mathrm{th}$, motivated by observations (Kennicut 1989), namely $f = 1$ for densities higher than $\rho_\mathrm{th}$ and $f = 0$ otherwise. \\
\noindent All assumptions lead to a self-regulating star formation because the growth of cold gas clouds is balanced by supernova feedback. \\
\noindent Finally, a prescription for galactic winds, according to \citet{springel03} is included. It is not a physically motivated model, but is trying phenomenologically to reproduce winds according to observational findings according to \citet{martin99}. It is assumed that the disc loses a certain amount of gas $\dot{M}_\mathrm{w}$ due to winds, which is proportional to the star formation rate, $\dot{M}_\mathrm{w} = \eta \dot{M}_\star$ with $\eta = 2$, according to the observations, and $\dot{M}_\star$ being the star formation rate of long-lived stars, namely with mass $M < 8\,\solarmass$. Additionally, a fixed fraction of the supernova energy of $\chi = 0.25$ is included in the wind, which contributes to the escape velocity of the wind from the disc. 

\subsection{Initial conditions of the model galaxies}

\begin{figure}[htb]
	\centering
	\subfigure[]{\includegraphics[width = \linewidth]{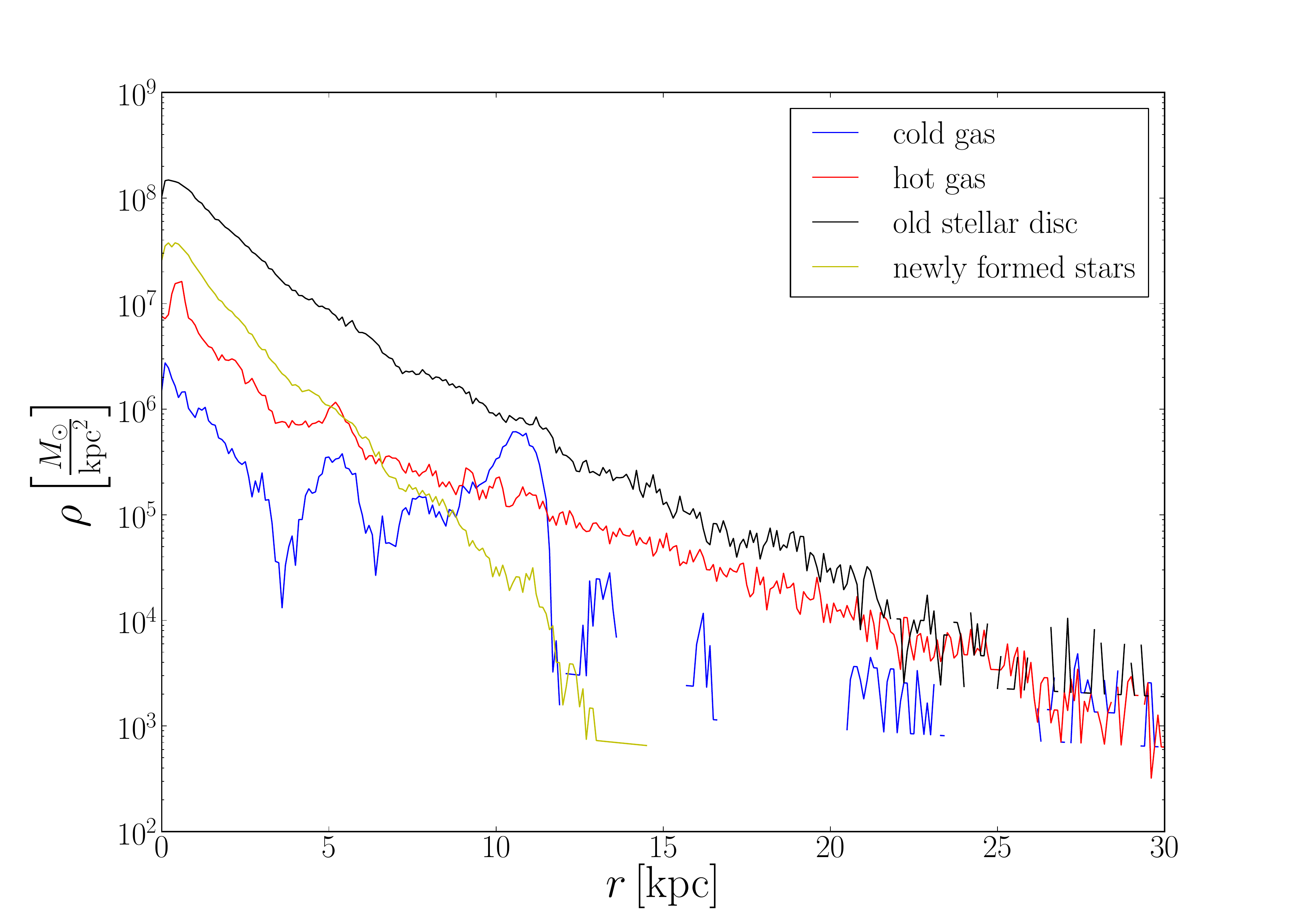} \label{fig:init_spiral_a}} \\
	\subfigure[]{\includegraphics[width = 0.8\linewidth]{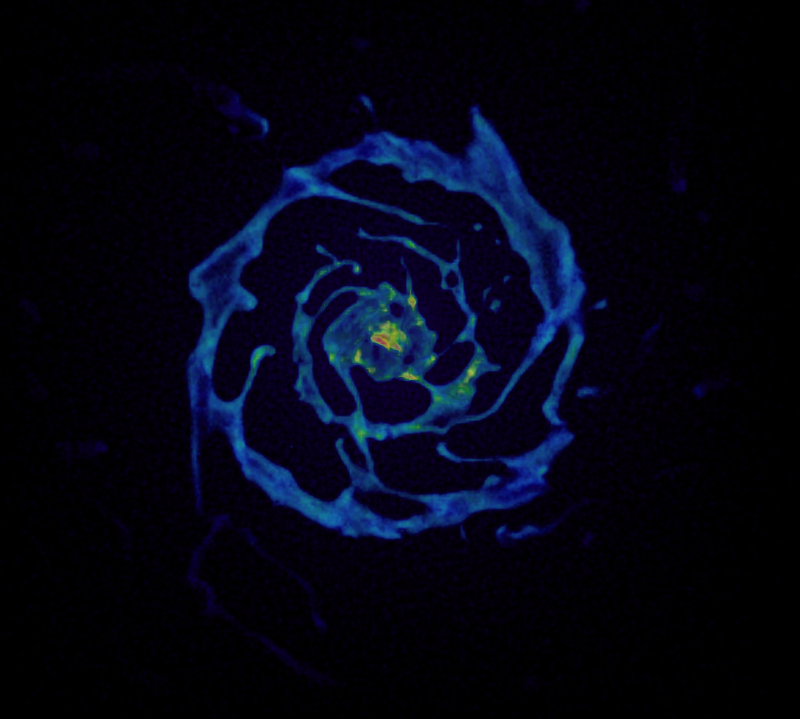} \label{fig:init_spiral_b}}
	\caption[]{(a) Radial density profile of model galaxy 1nb after 2 Gyr of evolution in the vacuum. We show two gas phases used in the simulation, named hot (ionized) and cold gas. In black, the old stellar disc is plotted and we show in yellow the newly formed stars within the 2 Gyr of evolution. Owing to the ring-like structure and the method of calculating the profile, it seems that some gas is missing at a radius of 15, 20, and 25 kpc. (b) Face-on view of the galaxy's gas component.  Already after 2 Gyr, overdensities in concentric rings are formed. Furthermore, a spiral structure can be seen.}
	\label{fig:init_spiral}
\end{figure}

The model galaxies were created with a disc galaxy generator developed by Volker Springel (details can be found in Springel et al. 2005) according to \citet{mmw98}. Three model galaxies with different bulges were created, all of them with a total mass of $1.08\times10^{12}\,\solarmass$. The gaseous disc has an initial mass of $2.38\times10^{9}\,\solarmass$. The bulge of the galaxies is modelled with collisionless particles such as the dark-matter halo and the stellar disc. Only the mass of the bulge is varied, the other mass fractions are unaltered to investigate the influence of the bulge. The halo circular velocity for all galaxies is $160\,\mathrm{km}\,\mathrm{s}^{-1}$ and the stellar disc-scale length varies from 2.8 to 3.3 kpc according to the different size of the stellar bulge. The mass resolution, number of particles and total mass used for the different model galaxies is shown in Table \ref{tab:initial_conditions}. Note that the additional mass of the bulge of galaxies 01b and 05b is removed from the dark-matter halo. Thus, all galaxies have the same total mass. Furthermore, galaxy 1nb has no bulge. \\
To avoid numerical artefacts, we evolved the model galaxies for 2 Gyr in isolation before they were used for further simulations of ram-pressure stripping. Because the initial condition generator simply generates a density and velocity distribution of the different components of a model galaxy, the evolution in vacuum, i.e. no surrounding gas, is necessary to gain self-regulated star formation. Furthermore, the gas distribution has settled to an equilibrium distribution with a spiral structure after 1 Gyr and a ring-like structure after 2 Gyr of evolution, as shown in Fig. \ref{fig:init_spiral_a} for model galaxy 1nb. The radial density profile and density distribution plot in Fig. \ref{fig:init_spiral_b} again shows model galaxy 1nb after 2 Gyr of isolated evolution. In blue and red we display the two gas phases (see Sect. \ref{sim_setup}), the hot ($> 10^4\,\mathrm{K}$) and cold ($< 10^4\,\mathrm{K}$) gas of the model galaxies. The several local maxima originate from the pattern of the spiral structure of the disc. Note that we consider only the gas particles instead of the smooth density distribution for the calculation of the density profiles. Combined with the considered annuli and the limited mass resolution of the particles, the density distribution seems to have empty areas at a radius of 15, 20, and 25 kpc. \\
\noindent During 2 Gyr of evolution, $4.2\times10^9\,\solarmass$ new stars have formed in model galaxy '1nb' with a current star formation rate of about $2.6\,\solarmass\,\mathrm{yr}^{-1}$. \\

\begin{table}[htb]
  \centering
  \begin{tabular}{rccc}
     & Galaxy '1nb' & Galaxy '01b' & Galaxy '05b' \\    
    \# of particles  \\
    \hline 
    DM halo \vline & $3\times10^{5}$ & $3\times10^{5}$ & $3\times10^{5}$ \\
    gaseous disc \vline & $2\times10^{5}$ & $2\times10^{5}$ & $2\times10^{5}$ \\
    stellar disc \vline & $2\times10^{5}$ & $2\times10^{5}$ & $2\times10^{5}$ \\
    stellar bulge \vline & 0 & $2\times10^{4}$ & $1\times10^{5}$ \\
    \hline \\
    mass resolution \\ $\left[ \solarmass / \mathrm{particle} \right]$ \\
    \hline
    DM halo \vline & $3.615\times10^6$ & $3.612\times10^6$ & $3.602\times10^6$ \\
    gaseous disc \vline & $1.191\times10^4$ & $1.191\times10^4$ & $1.191\times10^4$ \\
    stellar disc \vline & $3.572\times10^4$ & $3.572\times10^4$ & $3.572\times10^4$ \\
    stellar bulge \vline & - & $4.762\times10^4$ & $3.810\times10^4$ \\
    \hline \\
    total mass \\ $\left[ \solarmass \right]$ \\
    \hline
    DM halo \vline & $1.084\times10^{12}$ & $1.083\times10^{12}$ & $1.080\times10^{12}$ \\ 
    gaseous disc \vline & $2.381\times10^{9}$ & $2.381\times10^{9}$ & $2.381\times10^{9}$ \\ 
    stellar disc \vline & $7.143\times10^{9}$ & $7.143\times10^{9}$ & $7.143\times10^{9}$ \\ 
    stellar bulge \vline & $0$ & $9.524\times10^{8}$ & $3.809\times10^{9}$ \\
    \hline
  \end{tabular}
  \caption{Initial conditions of the three model galaxies with number of particles, mass resolution, and total mass of the different galaxy components. The total mass of all three model galaxies is the same. When including the bulges in model galaxies 01b and 05b, less mass is consigned to the dark-matter halo when initial conditions are generated.}
  \label{tab:initial_conditions}
\end{table}

\subsection{Simulating the ram pressure}

\begin{figure}[htb]
  \centering
  \includegraphics[width = 0.8\linewidth]{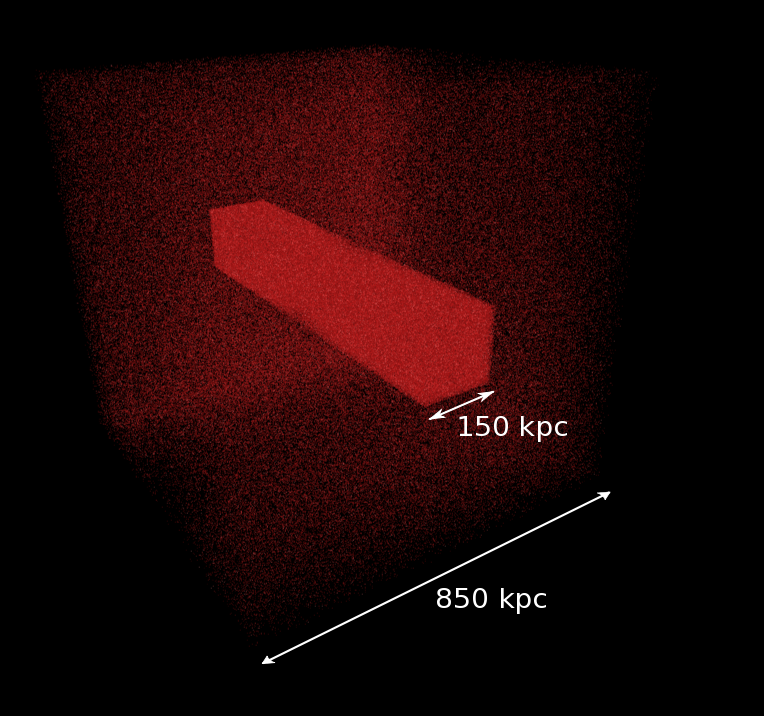}
  \caption{``Wind tunnel'' used to simulate the ram-pressure stripping. The low-resolution gas cube has a side length of 850 kpc. The high-resolution wind tunnel in the centre has a dimension of $150\times150\times850\,\mathrm{kpc}$.}
  \label{fig:wind_tunnel}
\end{figure}

To simulate a galaxy cluster environment, the model galaxies were exposed after 2 Gyr of evolution to an external wind, simulating the intra-cluster medium (ICM). For that purpose, a cuboid of gas particles was established, as depicted in Fig. \ref{fig:wind_tunnel}. In this case, contrary to other ram-pressure stripping simulations, the galaxies themselves are moving instead of the wind and the ICM is stationary, which leads to the same result. \\
\noindent The simulation domain was filled with gas particles of constant mass to achieve the desired density of the ICM. To avoid numerical artefacts due to the SPH scheme, the mass resolution of the particles representing the ICM was the same as that of the particles representing the ISM in the model galaxy. Because for this setup, GADGET-2 obliges cubic simulation domains, a box with 850 kpc side length was used. Filling this cube with particles of the given mass resolution would result in a huge amount of gas particles for the ICM and an unreasonable computing time. To avoid this problem, only in the centre a cuboid with a size of $150\times150\times850\,\mathrm{kpc}$ was filled with gas particles, which had the same resolution as the model galaxies. The surrounding of this cuboid consisted of a less resolved medium with a 100 times higher mass per particle, which kept the inner simulation domain, through which the model galaxies fly, stable. Additionally, periodic boundaries for the whole simulation domain were used.
Galaxies moving with a velocity of $1000\,\mathrm{km}\,\mathrm{s}^{-1}$ cross the whole simulation domain in about $0.83\,\mathrm{Gyr}$. The ICM does not change its density during this time by more than 4\%. \\
In the simulations, the model galaxies are moving with a constant velocity in a constant ambient medium. Since the ICM density is constant during the whole simulation, this can be thought of as galaxies orbiting a galaxy cluster, perceiving the same ram pressure throughout. The different configurations of ICM densities, model galaxies, and inclination angles are depicted in Table \ref{tab:simulations}. For all simulations, a relative velocity of $1000\,\mathrm{km}\,\mathrm{s}^{-1}$ was used. Since \citet{kapf09} found that the star formation rate depends much more strongly on the density of the ICM than on the relative velocity of a galaxy, we used only one velocity for the ram-pressure stripping simulations in this investigation. Furthermore, the ICM has a temperature of $10^7$ K.

\begin{table}[htb]
  \centering
  \begin{tabular}{c|ccc}
    
    \# & galaxy & ICM density $\left[ g\,cm^{-3} \right]$ & inclination angle $\left[ ^\circ \right]$ \\
    \hline \hline
    1 & 1nb & 0 & - \\
    2 & 1nb & $10^{-28}$ & 0 \\
    3 & 1nb & $10^{-28}$ & 22.5 \\
    4 & 1nb & $10^{-28}$ & 45 \\
    5 & 1nb & $10^{-28}$ & 67.5 \\
    6 & 1nb & $10^{-28}$ & 90 \\
    7 & 1nb & $10^{-27}$ & 0 \\
    8 & 1nb & $10^{-27}$ & 45 \\
    9 & 1nb & $10^{-27}$ & 90 \\
    \hline
    10 & 01b & 0 & - \\
    11 & 01b & $10^{-28}$ & 0 \\
    12 & 01b & $10^{-28}$ & 22.5 \\
    13 & 01b & $10^{-28}$ & 45 \\
    14 & 01b & $10^{-28}$ & 67.5 \\
    15 & 01b & $10^{-28}$ & 90 \\
    16 & 01b & $10^{-27}$ & 0 \\
    17 & 01b & $10^{-27}$ & 45 \\
    18 & 01b & $10^{-27}$ & 90 \\
    \hline
    19 & 05b & 0 & - \\
    20 & 05b & $10^{-28}$ & 0 \\
    21 & 05b & $10^{-28}$ & 22.5 \\
    22 & 05b & $10^{-28}$ & 45 \\
    23 & 05b & $10^{-28}$ & 67.5 \\
    24 & 05b & $10^{-28}$ & 90 \\
    25 & 05b & $10^{-27}$ & 0 \\
    26 & 05b & $10^{-27}$ & 45 \\
    27 & 05b & $10^{-27}$ & 90 \\
    \hline
    28 & 1nb (10.0 \%) & 0 & - \\
    29 & 1nb (17.5 \%) & 0 & - \\
    30 & 1nb (10.0 \%) & $10^{-28}$ & - \\
    31 & 1nb (17.5 \%) & $10^{-28}$ & - \\
    \hline
  \end{tabular}
  \caption{Properties of the different simulations. In simulations \#28--31, the disc gas
    fraction of the model galaxies was changed from the initial 25\% to 17.5\% and 10\%, respectively.}
  \label{tab:simulations}
\end{table}

\section{Results}

\subsection{The influence of a bulge and the inclination angle on the gaseous disc}

\begin{figure*}[htbp]
  \centering 
  \subfigure[]{\includegraphics[width=0.49\textwidth,trim=0 0.5cm 0 0]{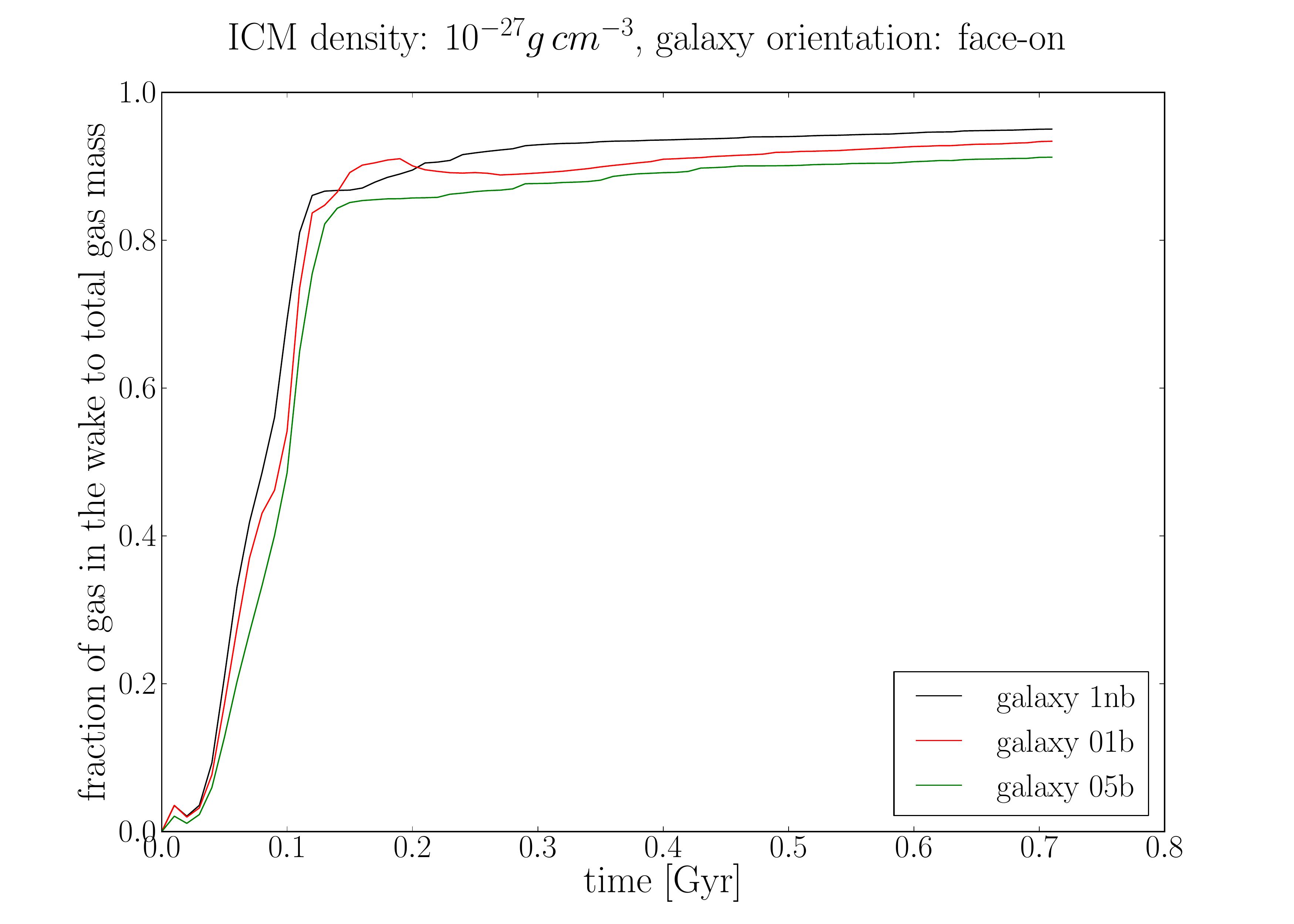}\label{fig:gas_distribution_a}}
  \subfigure[]{\includegraphics[width=0.49\textwidth,trim=0 0.5cm 0 0]{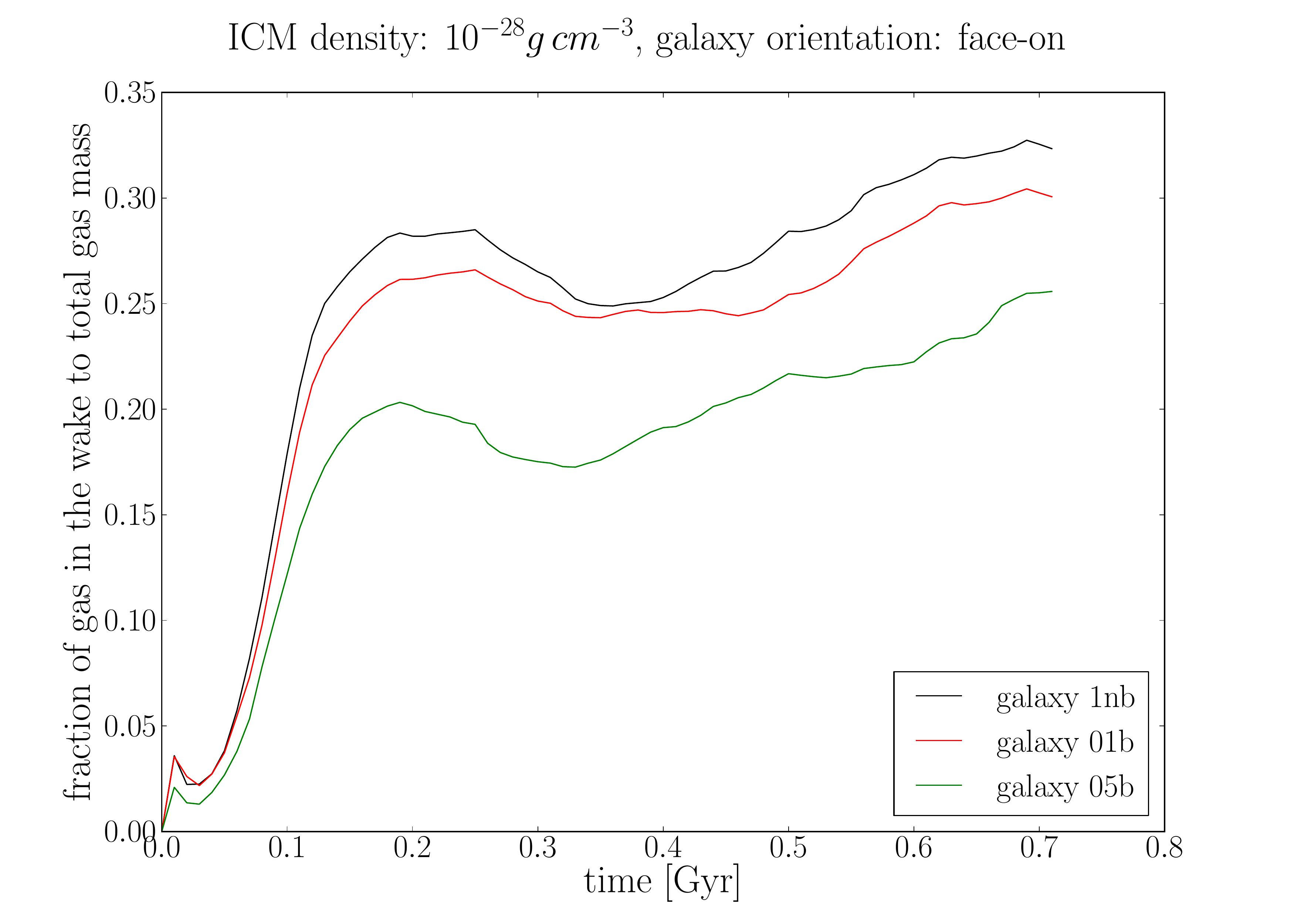}\label{fig:gas_distribution_b}}
  \subfigure[]{\includegraphics[width=0.49\textwidth,trim=0 0.5cm 0 0]{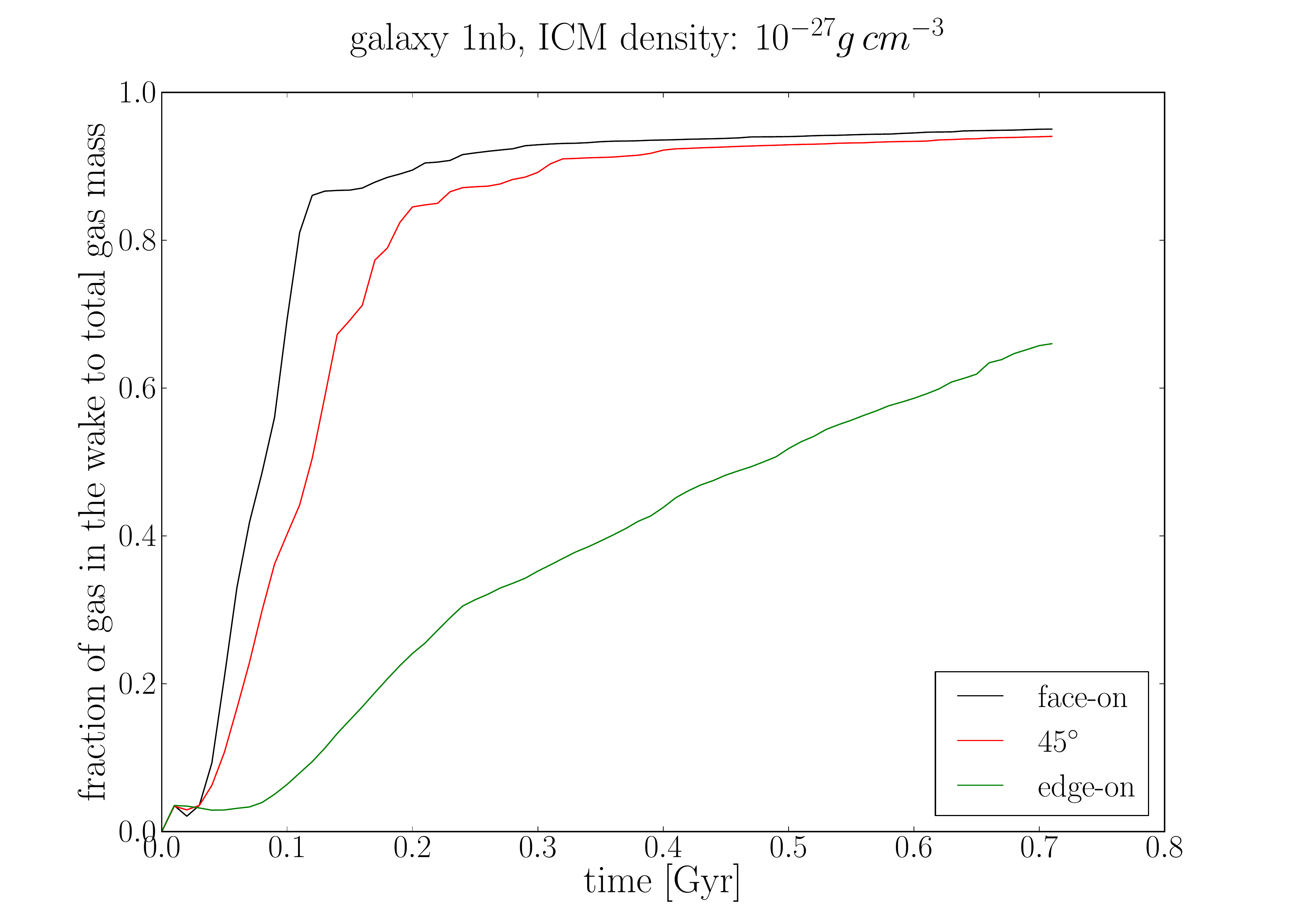}\label{fig:gas_distribution_c}}
  \subfigure[]{\includegraphics[width=0.49\textwidth,trim=0 0.5cm 0 0]{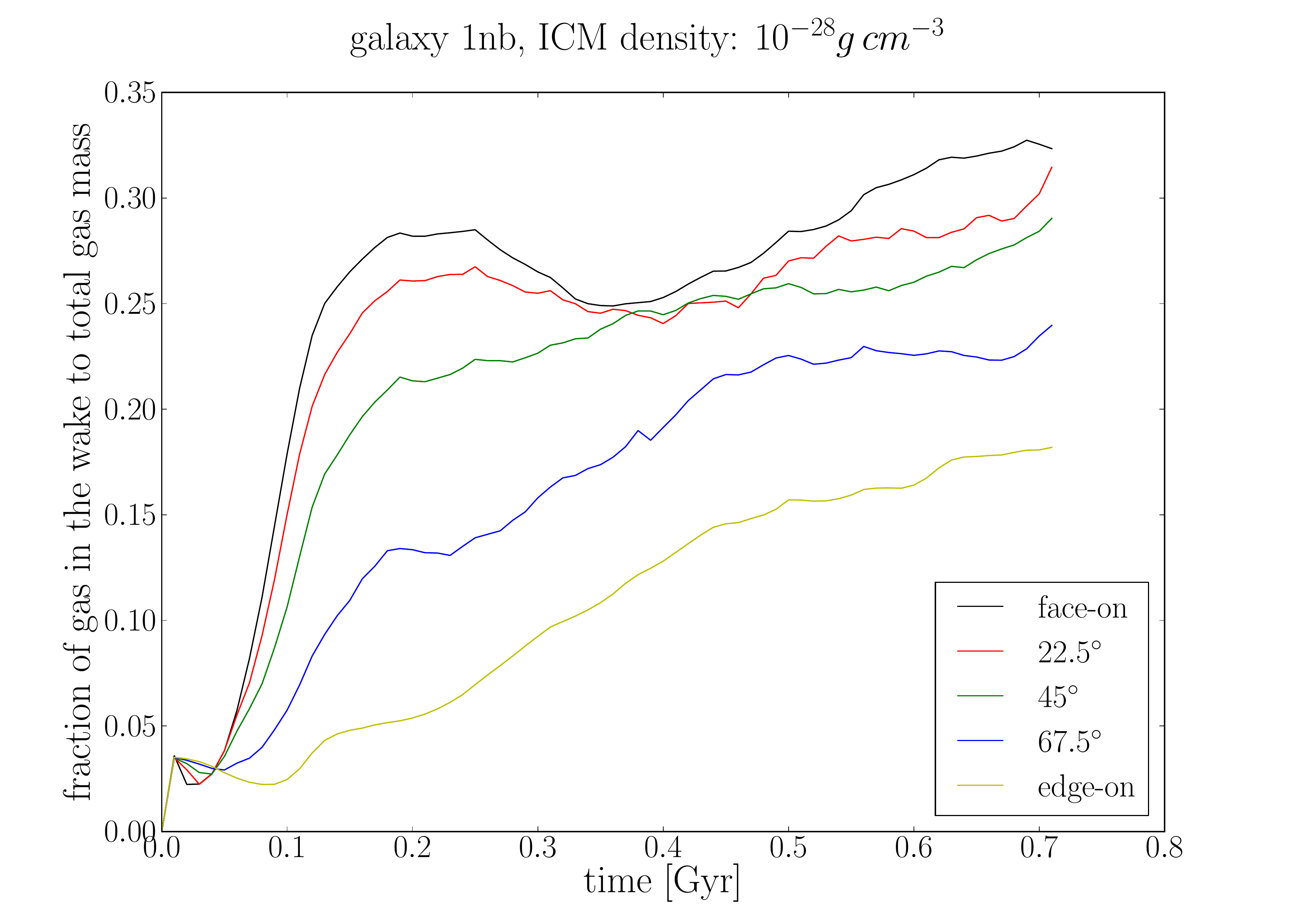}\label{fig:gas_distribution_d}}
  \caption{Fraction of the ISM, present in the wake of the galaxies, compared to the total amount of ISM is plotted versus simulation time. (a) and (c): high ram pressure, (b) and (d): low ram pressure. In (a) and (b), the behaviour of the different model galaxies is depicted to show the influence of a stellar bulge. In (c) and (d) we show the model galaxy 1nb, flying with different inclination angles with respect to the wind.}
  \label{fig:gas_distribution}
\end{figure*}

\begin{figure}[htbp]
  \centering
  \includegraphics[width=0.4\textwidth]{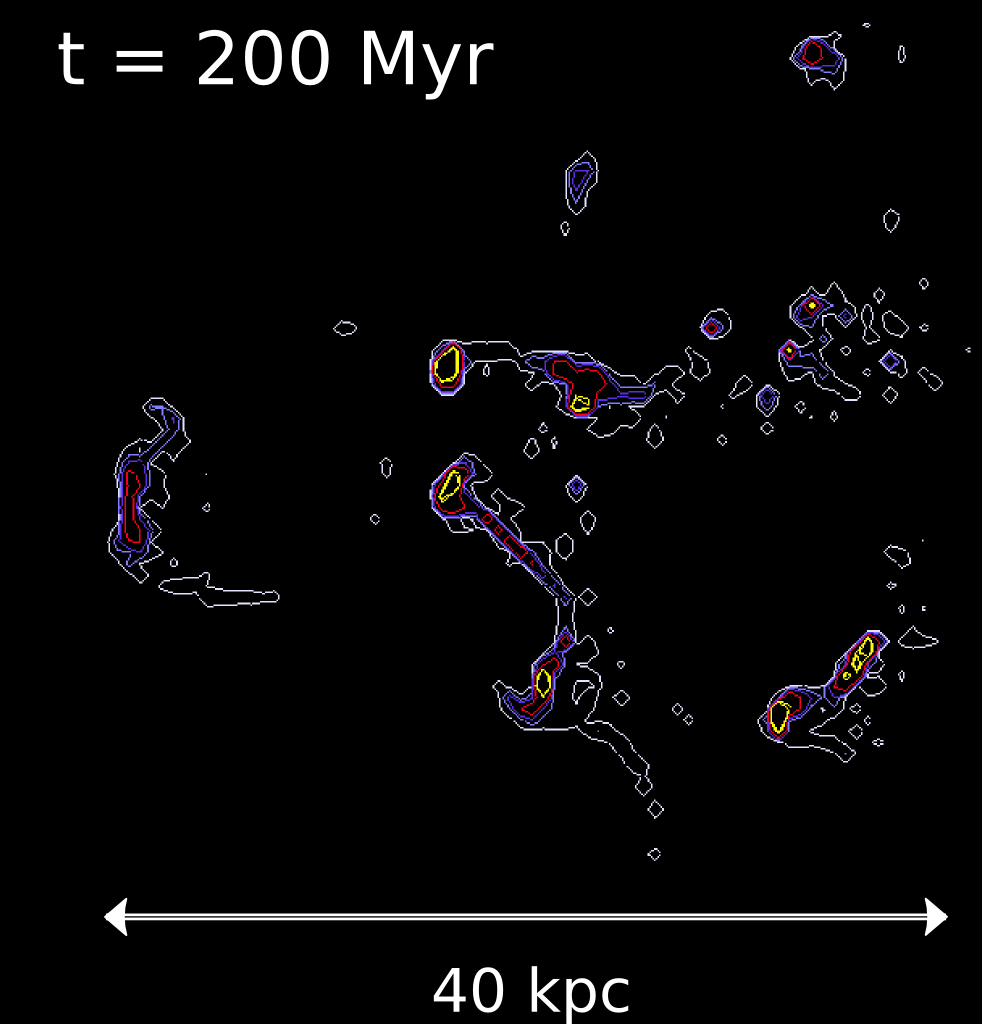}
  \caption{Surface density plot of the gaseous component of model galaxy 1nb after 200 Myr of acting ram pressure. The filamentary and knot-like structure in the wake, which already has a length of about 40 kpc, is obvious. The galaxy is located on the left-hand side where only a bunch of dense gas remains in the central region. In the picture, the galaxy moves towards the left.}
  \label{fig:gas_knot}
\end{figure}

First, we investigated the effect of the ram pressure on the gaseous disc. As expected and shown in recent simulations on ram-pressure stripping, at the beginning of the interaction of the galaxy with the ICM, the spiral arms are being compressed and the ISM density increases in the first 50 Myr of acting ram pressure. Moreover, the stripping of the gaseous disc begins in the outskirts of the galaxy. \\
\noindent In Fig. \ref{fig:gas_distribution}, we show the amount of stripped gas for the different simulations, and plot the fraction of gas in the wake of the galaxy to the total ISM gas mass versus simulation time. Obviously, when galaxies are flying face-on through the ICM, the gaseous disc is almost completely stripped after 100 Myr in the simulations with a high-density ICM of $10^{-27}\mathrm{g}\,\mathrm{cm}^{-3}$, and only a small amount of dense gas remains in the disc in the very centre. Owing to the rotation of the galaxy, filaments are forming in the stripped gas wake from the former spiral arms of the galaxy. After 200 Myr, very dense gaseous knots are forming in the wake. This scenario is also depicted for model galaxy 1nb in Fig. \ref{fig:gas_knot} where the gas surface-density is plotted. The gas tail containing these knots is 40 kpc long after 200 Myr. After 500 Myr of evolution the gas knots are ranging even to a distance of more than 300 kpc behind the disc. These self-gravitating gas knots are stable structures throughout nearly the whole simulation time although the mass of the knots is decreasing by 50--70 \% as depicted for one distinct gas knot in Fig. \ref{fig:knot_mass}. Moreover, the effect of a slipstream caused by the remaining gas disc can be observed. The filamentary structure is still present, albeit weakly. The gas density in these knots is comparable to the gas in the central region of the galaxy, reaching a value of $2.66\times10^{-24}\;\mathrm{g}\,\mathrm{cm}^{-3}$. \\
\noindent In the-low density ICM ($10^{-28}\mathrm{g}\,\mathrm{cm}^{-3}$) simulations, filaments in the gas tail can only be seen after 200 Myr, a small amount of gas is stripped (20--30 \%) and the main gaseous disc remains intact. After 500 Myr of acting ram pressure the filamentary structure reshapes to a few gaseous knots formed in the wake. Yet, the most dense regions are located at the centre of the galaxy. \\

\noindent 
The different model galaxies with distinct bulge masses behave differently when exposed to ram-pressure stripping, as can be seen in Fig. \ref{fig:gas_distribution_a} for a high-density and Fig. \ref{fig:gas_distribution_b} for a low-density ICM. Increasing the mass of the bulge, less gas is stripped, though the difference in the mass fraction is only a few percent for high ram pressure acting on a galaxy. If a low ram pressure is acting on the galaxy, the difference is more pronounced. In this special case, again the stripping is almost completed after 200 Myr. But with the low ram pressure acting, some part of the gas is falling back onto the disc, hence there is a local minimum at 300 Myr of simulation time. However, more gas is being stripped afterwards, which amounts to a total of a third of the gas. \\ 
\noindent
The impact of the inclination angle of a galaxy on the amount of stripped gas also depends on the ICM density, as depicted in Fig. \ref{fig:gas_distribution_c} and \ref{fig:gas_distribution_d} for model galaxy 1nb. As expected, with increasing inclination, a galaxy loses less gas. While for inclination angles below $45^\circ$ the stripping process is merely delayed, in the extreme case of galaxies flying edge-on through the ICM, only half of the gas is stripped in comparison to the face-on case. This influence of the inclination angle is the same for all model galaxies. \\

\subsection{The influence of a bulge and the inclination angle on the star formation rate}

\begin{figure*}[htbp]
  \centering 
  \subfigure[]{\includegraphics[width=0.49\textwidth,trim=0 0.5cm 0 0]{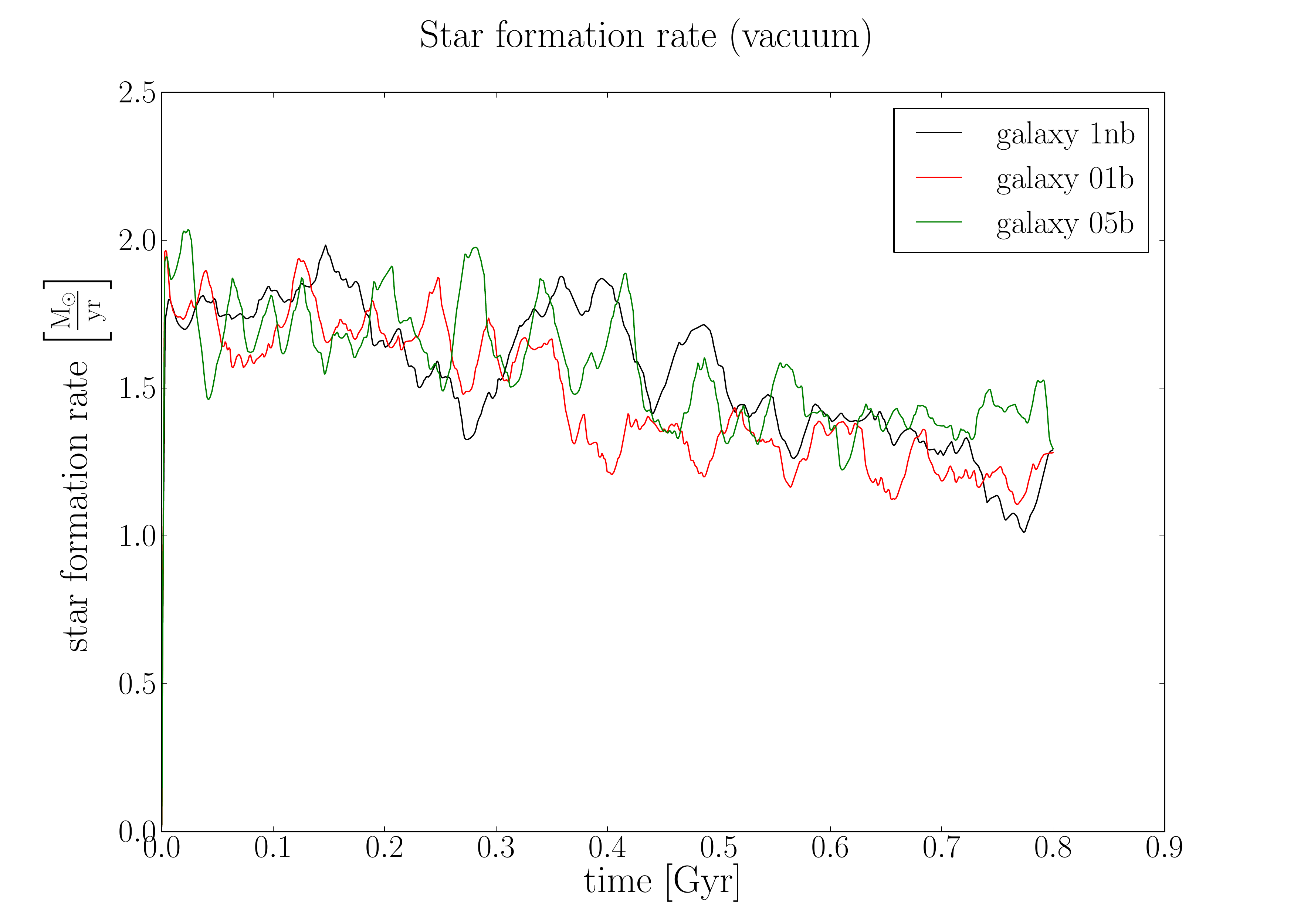}\label{fig:sfr_a}} \\
  \subfigure[]{\includegraphics[width=0.49\textwidth,trim=0 0.5cm 0 0]{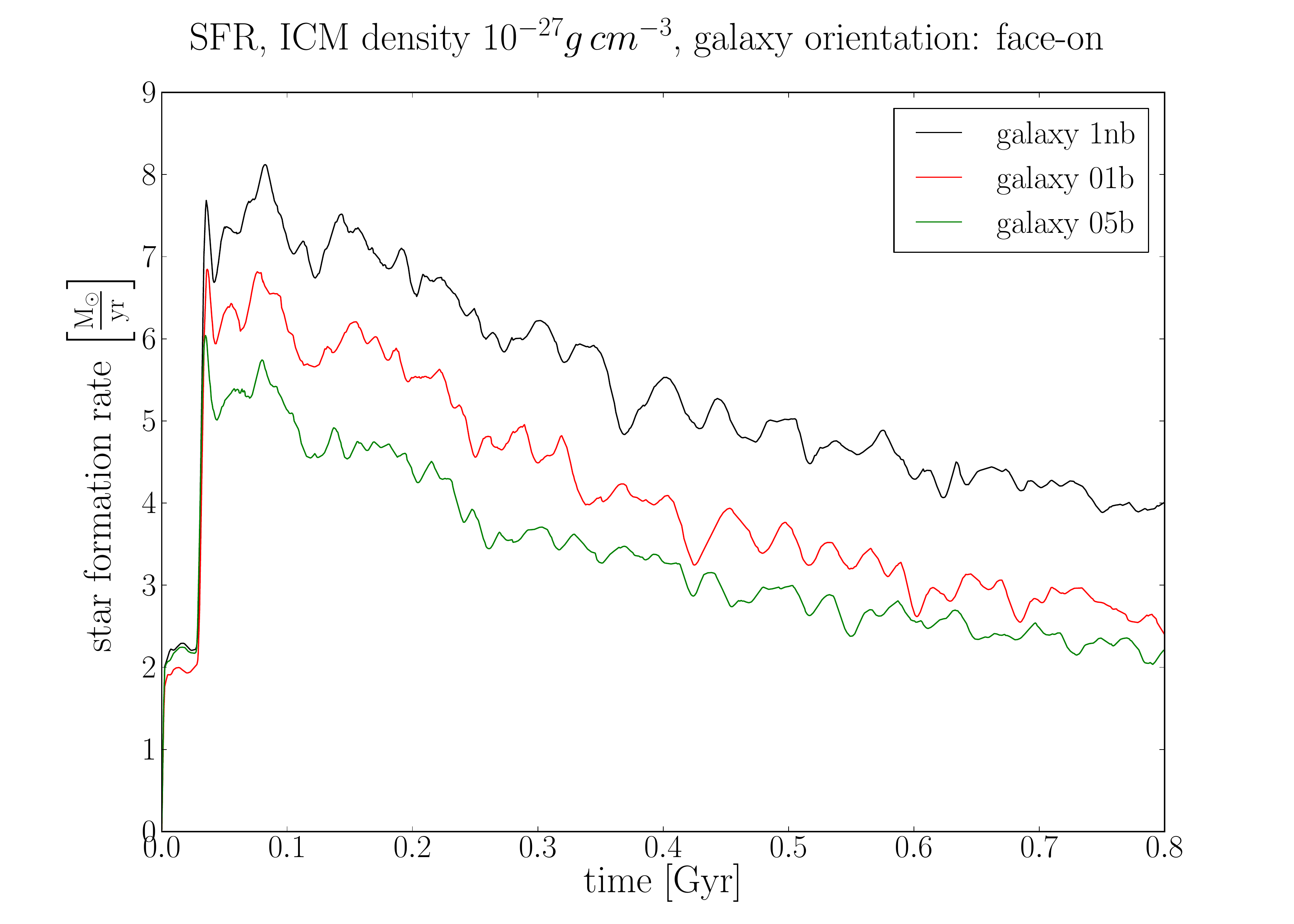}\label{fig:sfr_b}}
  \subfigure[]{\includegraphics[width=0.49\textwidth,trim=0 0.5cm 0 0]{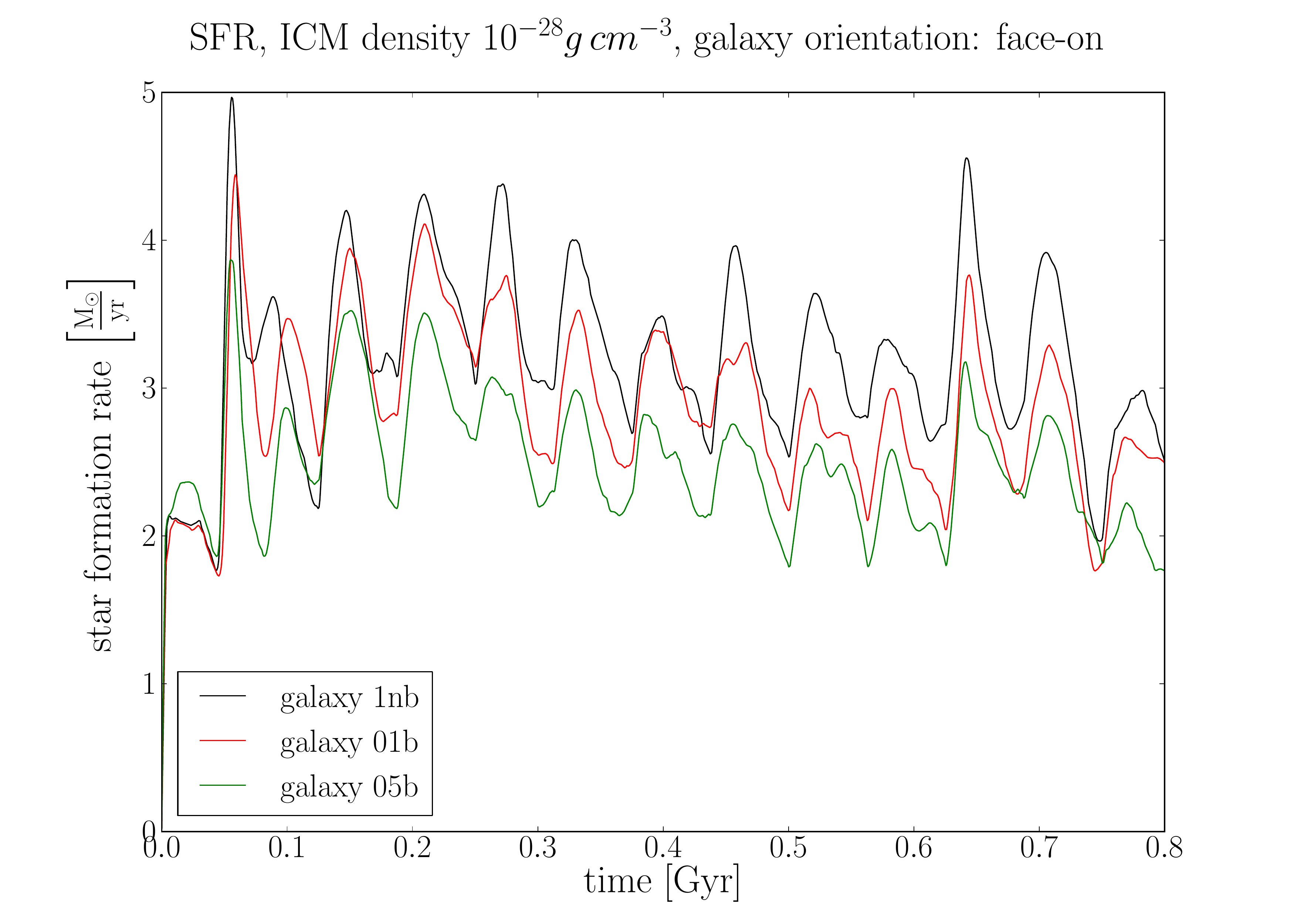}\label{fig:sfr_c}}
  \subfigure[]{\includegraphics[width=0.49\textwidth,trim=0 0.5cm 0 0]{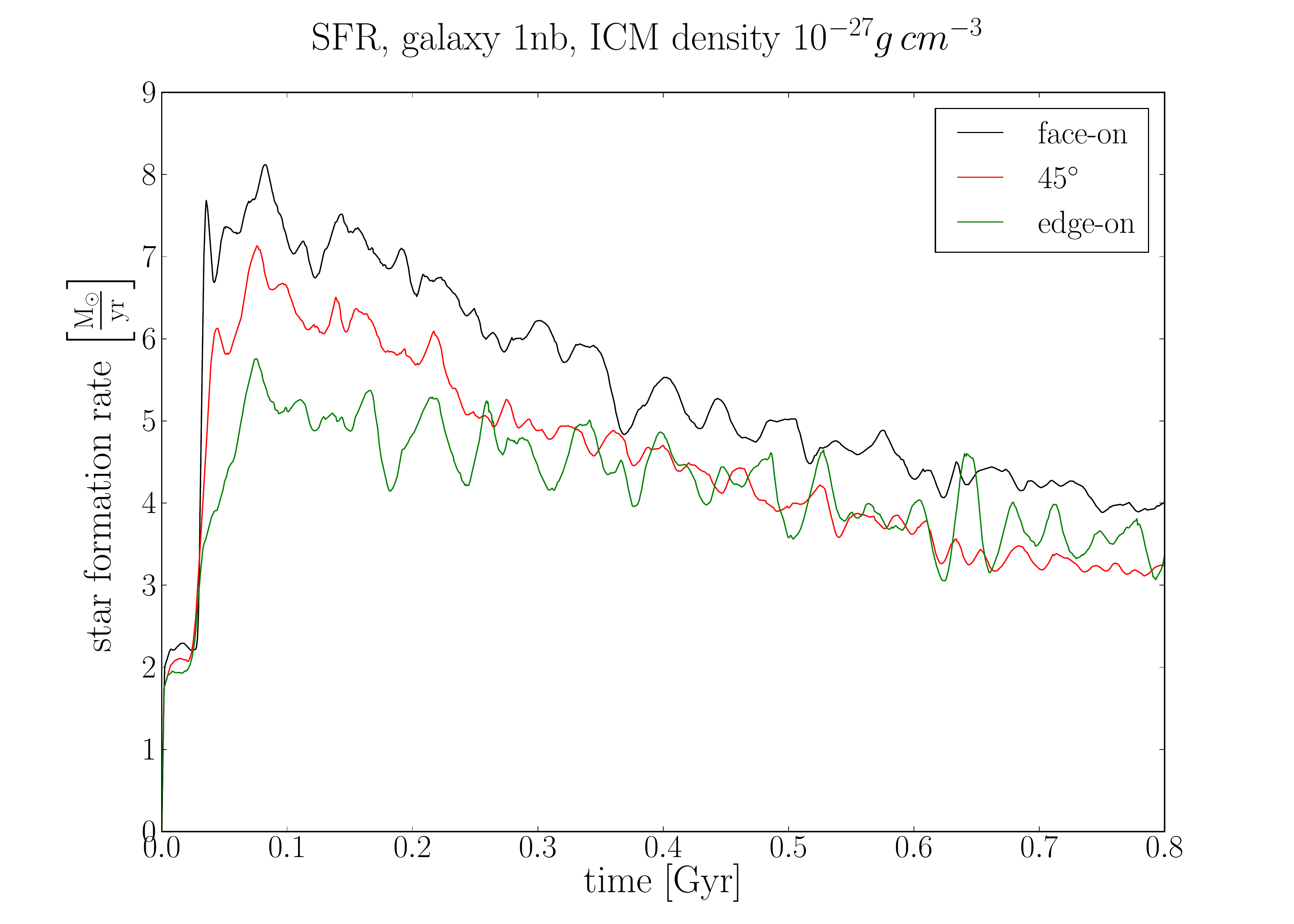}\label{fig:sfr_d}}
  \subfigure[]{\includegraphics[width=0.49\textwidth,trim=0 0.5cm 0 0]{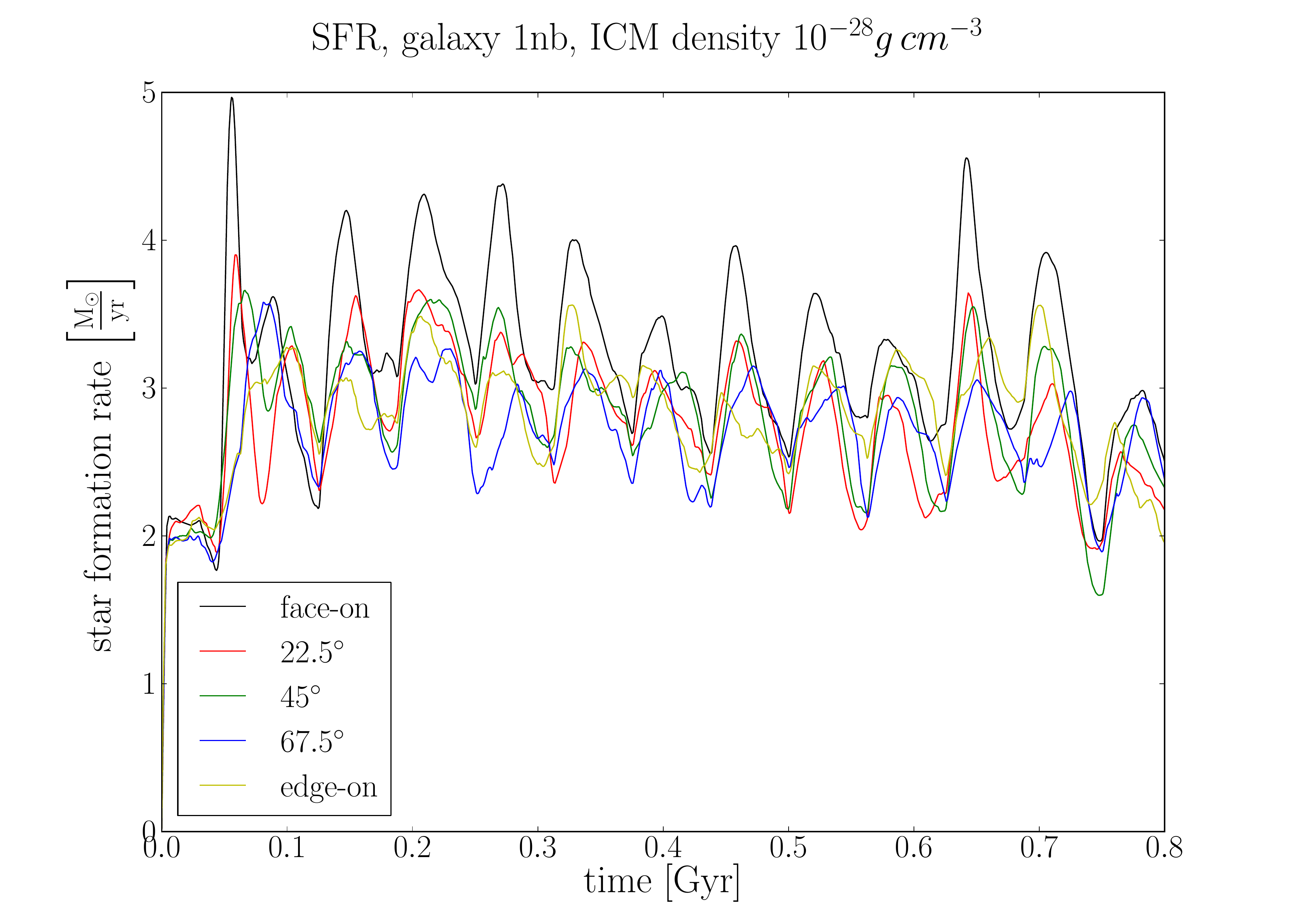}\label{fig:sfr_e}}
  \caption{Evolving star formation rates during the simulations. In Fig. (a) the star formation rates of the three model galaxies evolving in vacuum are shown, which are at about $1.5\,\solarmass\,\mathrm{yr}^{-1}$ for all galaxies. In Figs. (b) and (c) the star formation rates of the galaxies undergoing ram-pressure stripping are plotted, with an ICM density of (b) $10^{-28}\,\mathrm{g}\,\mathrm{cm}^{-3}$ and (c) $10^{-27}\,\mathrm{g}\,\mathrm{cm}^{-3}$. Already with a low ram pressure, the star formation is enhanced by a factor of two while with a high ram pressure the enhancement can reach a factor of 4 after the first interaction of the galaxies with the ICM. Clearly, a bulge suppresses the star formation rate. In Figs. (d) and (e) the star formation rate of model galaxy '1nb' undergoing ram-pressure stripping, flying with different inclination angles with respect to the ICM is shown. The star formation is suppressed if the galaxy is flying edge-on in case of high ram pressure. When a low ram pressure is acting on the galaxies, the inclination angle has almost no influence, the star formation rate is not affected.}
  \label{fig:sfr}
\end{figure*}

In Fig. \ref{fig:sfr}, the star formation rate of the whole simulation domain, including the wake, is plotted versus the simulation time for different scenarios. For comparison, the SFR for the three model galaxies in vacuum is plotted in Fig. \ref{fig:sfr_a}. \\
\noindent
If a high ram pressure (Fig. \ref{fig:sfr_b}, ICM density of $10^{-27}\;\mathrm{g}\,\mathrm{cm}^{-3}$, with a face-on orientation) is acting on the galaxies, the SFR is enhanced up to a factor of 4 compared to the same galaxies evolving in the field. Furthermore, a bulge decreases the star formation rate. For model galaxy 05b, the mean SFR is even 3 $\solarmass\,\mathrm{yr}^{-1}$ lower than that of model galaxy 1nb. \\
\noindent
When a low ram pressure is acting (Fig. \ref{fig:sfr_c}, ICM density of $10^{-28}\;\mathrm{g}\,\mathrm{cm}^{-3}$, with a face-on orientation), the star formation rate can be doubled. However, only a slight dependence on the bulge mass can be found, again showing a lower SFR with increasing bulge mass. But a periodic, sinusoidal change in the star formation rate can be observed because of the properties of the star formation recipe: in a periodic manner, the gas cools down and is compressed, hence new stars are formed. Consequently, the surrounding of the newly formed stars is heated up again due to the stellar feedback, which also decreases the density and hence lowers the SFR. Then, this process starts again. \\
\noindent
When the inclination angle of a galaxy relative to the movement is changed, the influence on the SFR is not very pronounced. In Figs. \ref{fig:sfr_d} and \ref{fig:sfr_e}, the SFR of model galaxy 1nb is shown for a high and low ram pressure, respectively, and with different inclination angles. With a low-density ICM and hence a low ram pressure, the sinusoidal shape of the star formation rate can be seen again and only a slight trend of decreasing star formation with increasing inclination angle can be observed. Obviously, the highest SFR can be seen in the face-on case, which is almost $0.5\,\solarmass$ higher than when the galaxy is tilted. If the inclination angle is increased, this effect becomes less pronounced. If there is high ram pressure, as already seen in the face-on case, the SFR is more enhanced for all inclination angles than in the case of low ram pressure. Again, the highest SFR can be seen in the face-on case, with decreasing SFR as the inclination angle increases. Interestingly, for a $45^\circ$ angle compared to the edge-on case, the SFR is higher in the beginning of the simulation, while it is lower at the end, resulting in almost the same amount of new stars being produced.

\subsection{The influence of the disc gas fraction on star formation rate and ram-pressure stripping}

\begin{figure}[htb]
  \centering
  \subfigure[]{\includegraphics[width = 1.0\linewidth]{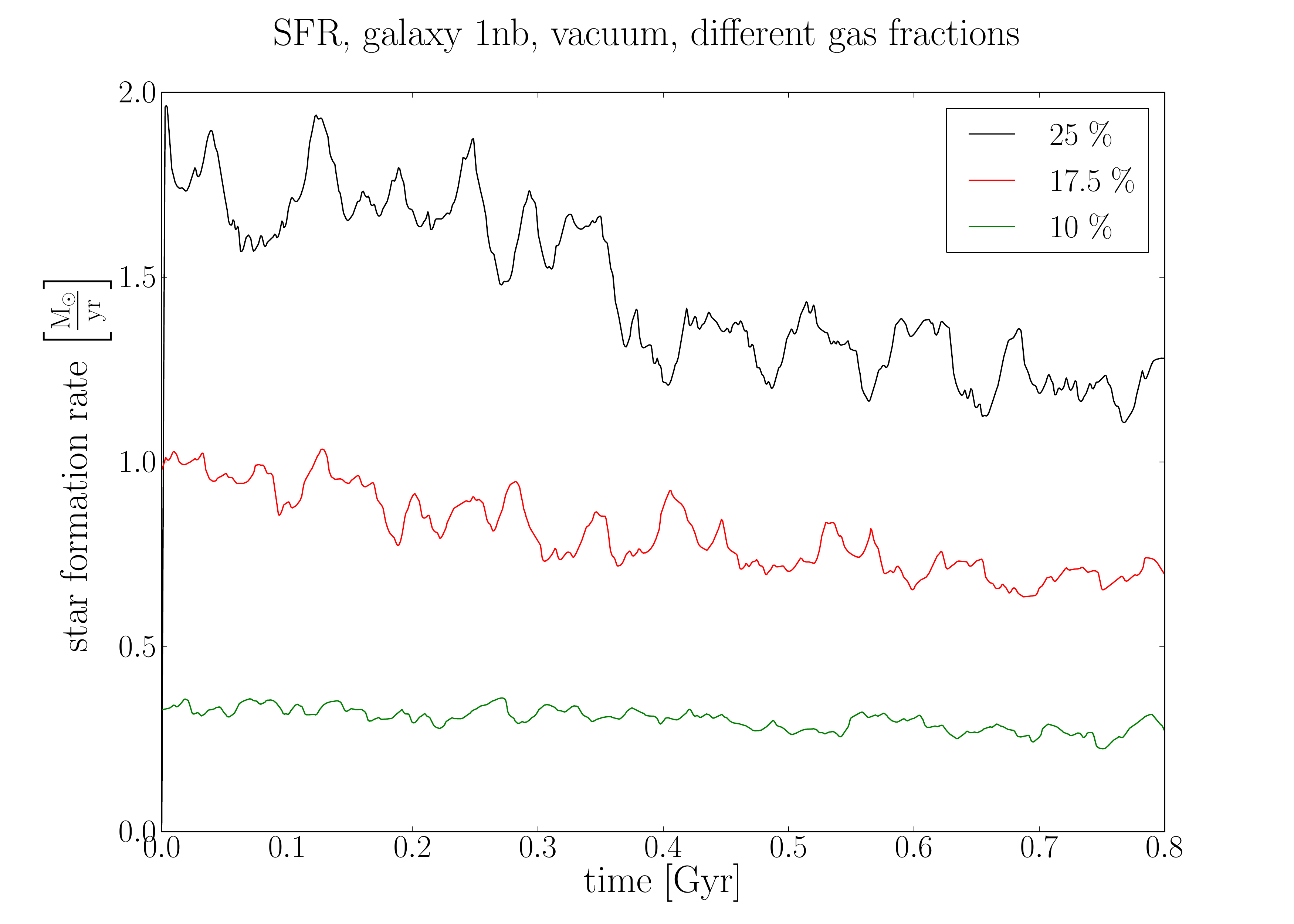}\label{fig:sfr_gasfrac_a}}
  \subfigure[]{\includegraphics[width = 1.0\linewidth]{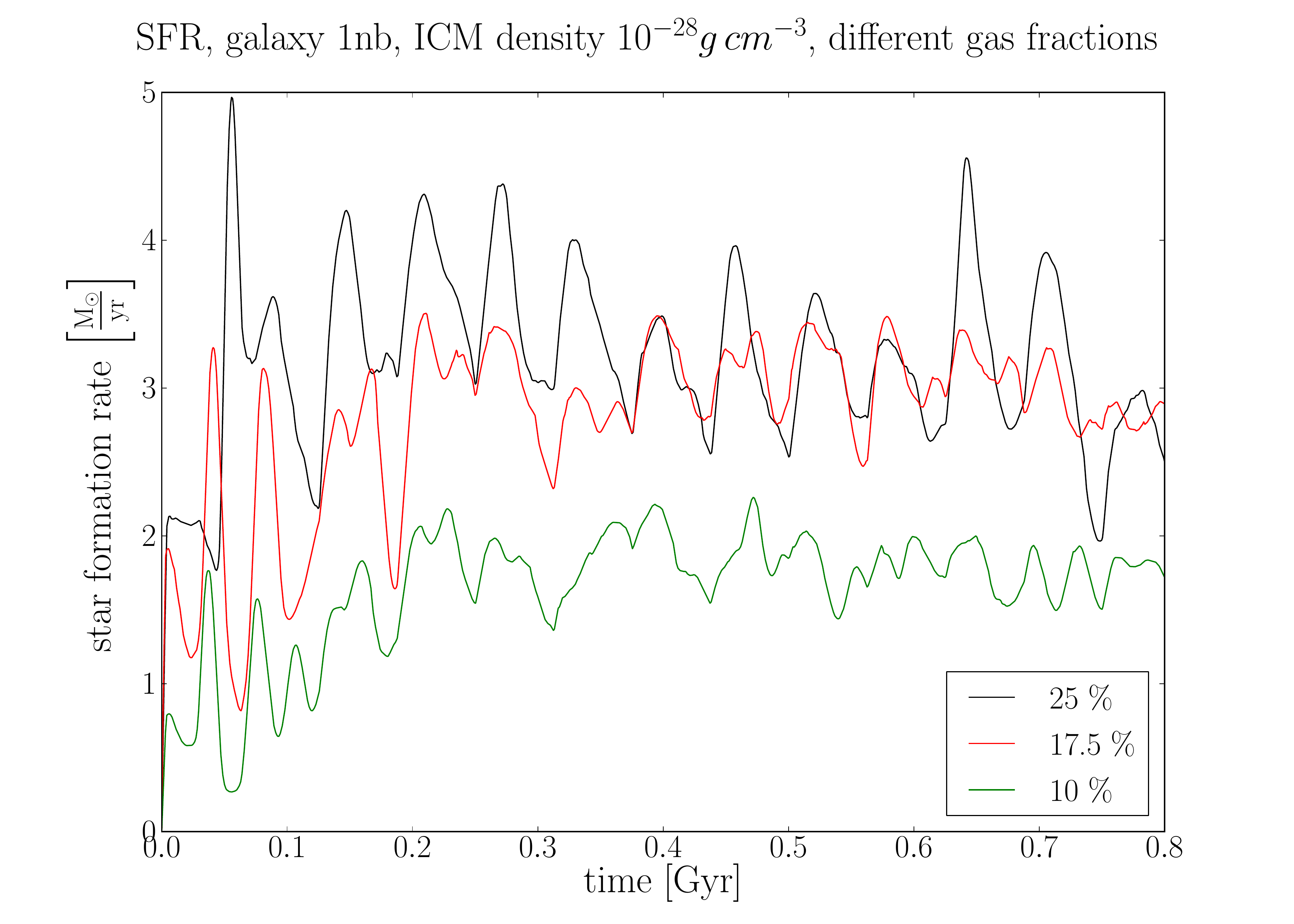}\label{fig:sfr_gasfrac_b}}
  \caption{Star formation rates in simulations \#1 and \#28--30 are plotted. In (a), SFR of galaxy 1nb with different gas fractions in the disc are plotted, 
    evolving in vacuum. In (b), SFRs of the same three model galaxies undergoing ram-pressure stripping are displayed. As expected, a lower gas mass fraction in the disc implies 
    a lower star formation rate. In the ram-pressure stripping case, the star formation is enhanced in all cases, but with increasing gas mass fraction, the effect is less 
    pronounced.}
  \label{fig:sfr_gasfrac}
\end{figure}

\noindent 
In addition to considering the differences in SFR of a galaxy and the amount of stripped gas in a ram-pressure stripping
scenario, the influence of the gas mass fraction on the disc needs to be scrutinised as well, which we did for model galaxy 1nb. In Fig. \ref{fig:sfr_gasfrac}, the evolving SFRs 
of simulations \#1 and \#28--30 are displayed. In Fig. \ref{fig:sfr_gasfrac_a}, the SFRs of model galaxy 1nb evolving in vacuum are shown, with gas mass fractions of 25\%, 17.5\%, and 10\%
respectively. In Fig. \ref{fig:sfr_gasfrac_b}, the SFRs of the same model galaxies undergoing ram-pressure stripping in an ICM with density $10^{-28}\,\mathrm{g}\,\mathrm{cm}^{-3}$ are displayed. Obviously, increasing the gas mass fraction in the disc leads to a higher star formation rate both when galaxies evolve in vacuum 
and when they undergo ram-pressure stripping. But with decreasing gas mass fraction, the SFR is more enhanced by ram-pressure. \\
\noindent
The fractional amount of stripped gas is independent of the gas mass fraction in the disc. Because the whole disc mass, including 
the stars, remains the same for the model galaxies, the restoring force according to the Gunn \& Gott criterion also stays the same. Hence, 
these statements are also valid for the model galaxies with different bulge sizes. \\

\subsection{Distribution of the ISM and newly formed stars}
\label{sec:distribution}

\begin{figure*}[htbp]
  \centering
  \includegraphics[width=\textwidth]{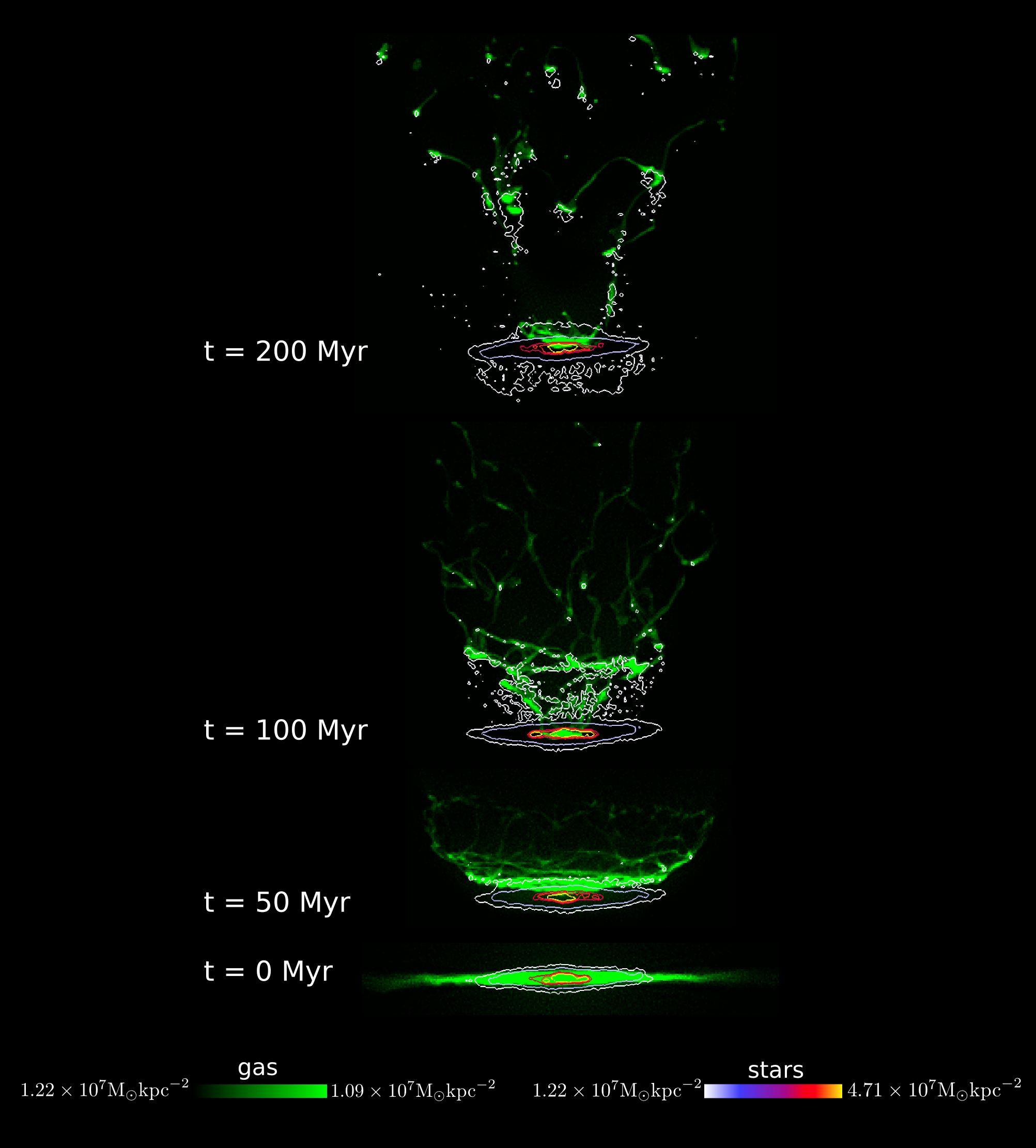}
  \caption{Surface density of the ISM (green) and the newly formed stars (isolines)
    are shown for a ram-pressure stripping scenario of model galaxy 1nb in an ICM with a density of $10^{-27}\,\mathrm{g}\,\mathrm{cm^{-3}}$. 
    Note the colourbars for both distributions. Four different time steps are shown.
    New stars in the wake are formed in the dense gas knots. After 200 Myr many stars
    are present also in front of the disc. The stars formed in the wake are gravitationally attracted
    by the disc and due to the collisionless dynamics, these stars from the wake are falling
    through the disc. Furthermore, with the initial onset of ram pressure, the gas disc is pushed 
    back from the stellar disc, as can be seen in the timestep after 50 Myr of evolution in the ICM.
    }
  \label{fig:surface}
\end{figure*}

\begin{figure}[htbp]
  \centering
  \subfigure[]{\includegraphics[width=\linewidth]{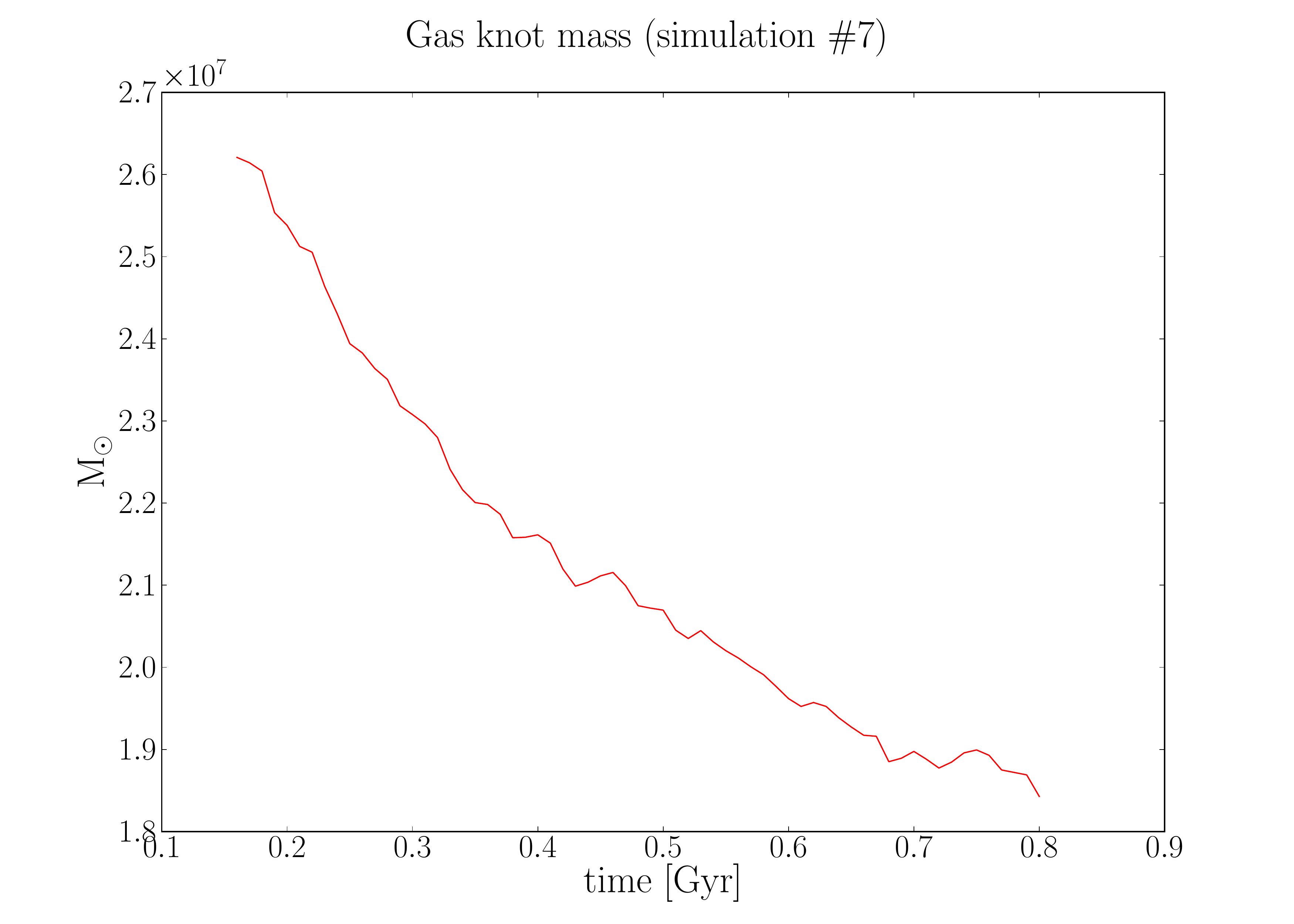}\label{fig:knot_mass_a}}
  \subfigure[]{\includegraphics[width=\linewidth]{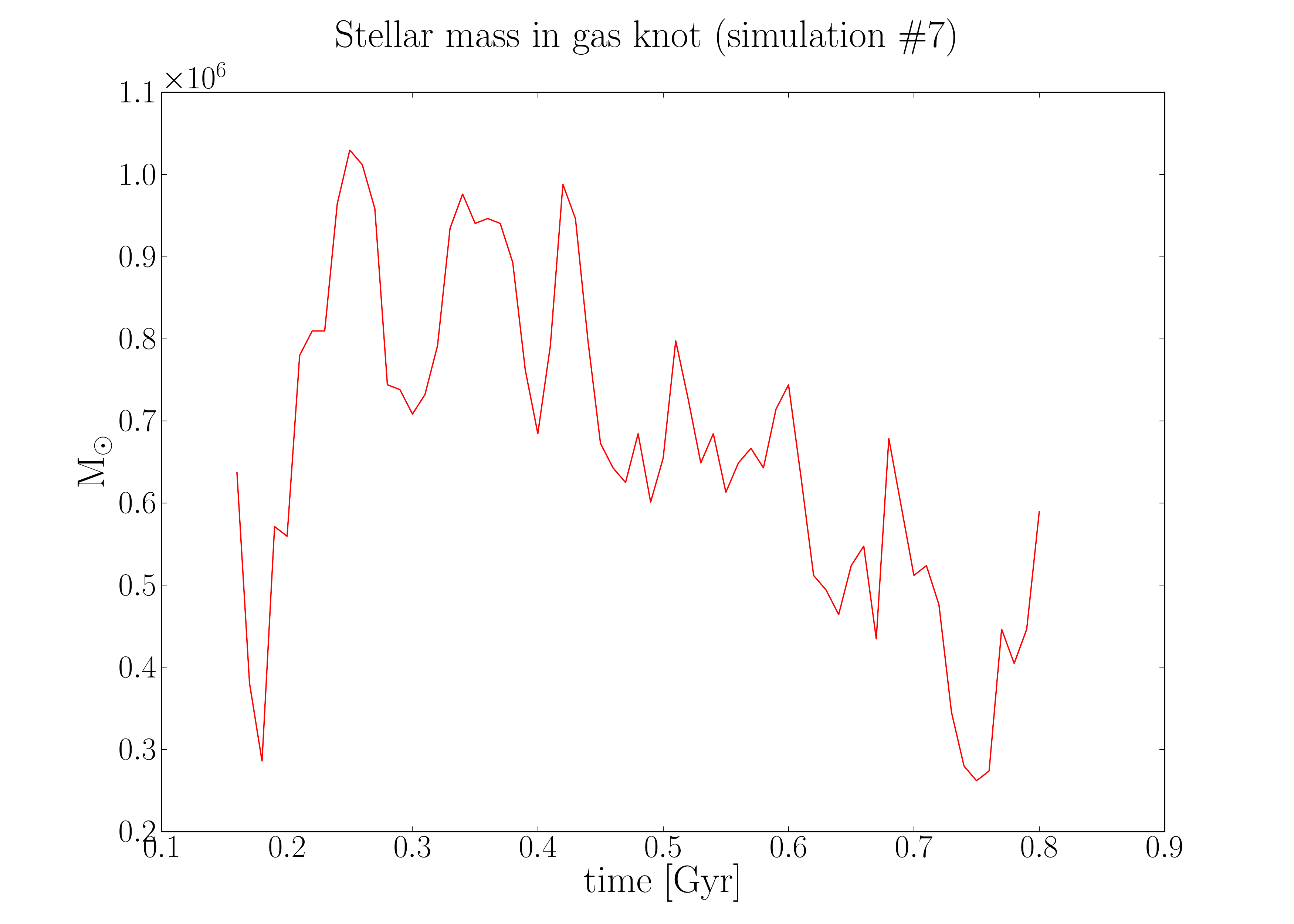}\label{fig:knot_mass_b}}
  \caption{Mass of a distinctly chosen gas knot in the wake of ram-pressure stripped galaxy 1nb in simulation \#7 is shown with respect to
    simulation time. The gas clump is formed after 160 Myr of the onset of the ram pressure. Until the end of the simulation, the gas knot loses half
    of its mass due to star formation as shown in (a). The mass of newly formed stars, present in the gas knot, is displayed in (b).}
  \label{fig:knot_mass}
\end{figure}

As shown before, when the ram pressure acting on a galaxy is high, almost all of its gas can be stripped. Nevertheless, the system of galaxy plus stripped gas still shows a high SFR. Obviously, the stars are formed in the stripped gas wake of the galaxy. \\
In Fig. \ref{fig:surface}, four different timesteps of simulation \#7 are shown with the gas and newly formed stars' surface density. Evidently, after 200 Myr, several dense gas knots are present in the wake with aggregated stars that were formed in these dense gas knots. Since the stars do not feel the ram pressure any more and are attracted by the galaxy's potential, they are falling towards and through the disc.  Hence, stars are present also in front of the disc, as depicted in the timestep at 200 Myr. \\
\noindent
The first gas knots are formed after about 150 Myr of interaction with a dense ICM or rather when a high ram pressure is acting on a galaxy. The knots are present throughout the whole simulation for a total interaction time of the galaxy with the ICM  of 800 Myr. Nevertheless, the gas knots are losing mass, as shown in Fig. \ref{fig:knot_mass_a} for a gas knot taken from simulation \#7. The mass loss is only due to star formation, thus the lost gas mass is transformed into new stars. The stellar feedback is not able to destroy the gas knots or to contribute to a significant outflow of the gas from the knots. Hence, cooling processes, self gravity and the external pressure from the ICM keep the gas knots stable over time 
and lead to an SFR within these knots throughout the whole simulation in the range of 0.025 to 0.01 $\solarmass\,\mathrm{yr}^{-1}$. 
In Fig. \ref{fig:knot_mass_b}, the mass of newly formed stars
in the distinct gas knot is shown throughout the simulation. Up to $10^6\,\solarmass$ of newly formed stars can exist within the gas knot. 
Continuously, stars are falling back onto the galactic disc, creating streams of stars in the wake of the galaxy, as shown in timestep $t = 200\,\mathrm{Myr}$ in Fig.
\ref{fig:surface}. \\
\noindent
The amount of stars formed in the wake is a direct consequence of the amount of stripped gas, which differs for distinct bulges and different inclination angles of a galaxy and of course for different ram pressure. With a low ram pressure, very few stars with a
cumulative mass of $\approx 4 \times 10^6\, \solarmass$ are present in the wake after 500 Myr of interaction with the ICM. With increasing
inclination angle of a galaxy, also fewer stars are present in the wake as less gas is stripped. Consequently, also the presence of a bulge leads to less star formation in the wake, as expected.

\section{Comparison with observations and other ram-pressure stripping simulations}

In this work, an SPH scheme as in \citet{kronberger06, kronberger07} and \citet{kapf09} was used. In comparison to other ram-pressure stripping simulations that use a grid code \citep{roediger06, roediger07}, the behaviour is  quite different. In the grid codes, Kelvin-Helmholtz or Rayleigh-Taylor instabilities are much better resolved. In this case, fewer or even no gas clumps are formed. Different approaches for solving the hydrodynamic equations such as modified SPH schemes \citep[e.g.][]{hess10, read10} or hybrid eulerian-lagrangian codes as in \citet{springel10} also show less clumpiness in a galaxy's wake. However, self-gravity, cooling processes, and the external pressure from the ICM can indeed contribute 
to the stability of the gas clumps in a galaxy's wake, which then lead to star formation occurring outside of galaxies. Furthermore, the initial spiral structure of the gaseous component in the model galaxies, which are in a first moment compressed, lead to the filamentary structure and the final formation of the gas clumps. \\ 
\noindent
Galaxies undergoing ram-pressure stripping, as shown in our simulations, have been observed e.g. in the Virgo Cluster. The VIVA survey (VLA imaging of 
atomic gas in the Virgo Cluster) shows galaxies NGC4402 and NGC4522 \citep{crowl05, crowl06}, which are prominent examples for interactions of the interstellar 
with the intra-cluster medium. For these cases, Crowl and collaborators showed the possibility that galaxies can lose almost all of their gas in the case of high ram-pressure,
as showed also apparently in the simulations. Furthermore, it can be seen that ram-pressure stripping can trigger a starburst with a subsequent cessation of the SFR,
especially if the galaxy loses all of its gas. Such a strong ram-pressure effect can not only be found in central regions of a galaxy cluster but also in regions well
outside the cluster core, which affirms the validity of the orbiting scenario in our simulations. \\
\noindent
Another aspect is the transformation of spiral galaxies into S0 or early-type galaxies. \citet{tran05} found by studying galaxies in MS2053 at 
redshift $z = 0.6$ that the Butcher-Oemler effect is linked to galaxy infall. In this case, as expected and seen in the simulations, ram-pressure
stripping also plays an important role. Furthermore, \citet{wolf09} found in the context of the STAGES survey that the star formation rate in the observed cluster
is clearly lower than in the field. Blue spirals in the field might be transformed first into red spirals and eventually into S0s 
in the cluster centre during the infall. Again, ram-pressure stripping, as shown in the simulations, might play an important role. \\
\noindent
The influence of the inclination angle of a galaxy's movement with respect to the ICM on the amount of stripped gas, which we have found, is
confirmed by \citet{boesch12}. The galaxies in the Abell 901/902 supercluster show a clear correlation in the sense that a more face-on infall of a galaxy leads to a smaller 
gas disc with respect to the stellar disc, indicating a higher amount of stripped gas. \\
\noindent
Another prominent example is the galaxy IC 3418 in the Virgo cluster, observed with GALEX \citep{martin05}. It has been found that this galaxy has an ultraviolet tail of bright knots and diffuse emission with a length of about 17 kpc. Near ultraviolet, far ultraviolet and $\mathrm{H}\alpha$ emissions indicate that star formation is ongoing \citep{gavazzi06}. \citet{hester10} suggested, that IC 3418 must have been ram-pressure stripped in order to lose that amount of gas. As in the simulations, these authors claim that the gas clumps must have been formed in the stripped wake, aided by turbulence. Furthermore, the suggested star formation is taking place in these dense gas knots as well. \\
\noindent
More evidence for gas knots forming in the wake of ram-pressure stripped galaxies is provided by observed 'fireballs' in the galaxy RB199 
in the Coma cluster by \citet{yoshida08}. This galaxy shows a highly disturbed morphology that indicates a galaxy-galaxy merger. Interestingly, 
the galaxy also features filaments and bright blue knots at the ends of these filaments, which strongly emit $\mathrm{H}\alpha$ and ultraviolet radiation,
indicating ongoing star formation. \citet{yoshida08} concluded, that ram-pressure stripping caused these knots to form.
The masses of these knots are $\sim 10^6-10^7\,\solarmass$, corresponding to the faintest dwarf spheroidals in the Virgo cluster (ibid.) 
and agreeing well with the masses found in the gas knots in our simulations, discussed in Sect. \ref{sec:distribution}. 
\citet{yoshida08} found a typical SFR of around $10^{-3}\,\solarmass\,\mathrm{yr}^{-1}$,
which is lower than in our simulation. But because the knots of RB199 consist mainly of young stars formed there, having depleted the gas reservoir, they should be
in a later evolutional state than in our simulations. Because RB199 is in a post-starburst phase, introduced by the merger event, another possibility
is that this galaxy already consumed a lot of its gas before the onset of ram-pressure stripping, which would explain the discrepancy in the observed and
simulated star formation rates. \\
Moreover, the filaments, consisting of older stars, are possibly left behind during the stripping, because they do not feel the ram pressure
and are attracted by the gravitation of RB199. This can be well explained by our simulations, where we also found a fraction of 
stars falling back to the disc. \\
\noindent
We aim in the near future to compare observations based on our own data. In this project, we will use deep narrow- and broad-band imaging with VLT/FORS2 to trace in-situ star formation in the gas wakes of two disc
galaxies in the Virgo cluster.

\section{Discussion and conclusion}

We carried out numerical studies on the effect of ram-pressure stripping and especially
the influence of a galaxy's bulge. Thirty-one different high-resolution N-body/SPH
simulations were conducted. The size of the galaxy bulge was varied and we used
two different ICM densities of $10^{-28}\,\mathrm{g}\,\mathrm{cm}^{-3}$ and $10^{-27}\,\mathrm{g}\,\mathrm{cm}^{-3}$. In each case, the
relative velocity of the galaxies was $1000\,\mathrm{km}\,\mathrm{s}^{-1}$ and the temperature of the ICM was $10^7$ K. \\

- The size and mass of the bulge of the model galaxies were varied. While the total mass of all galaxies remained the same, 
mass was shifted from the dark-matter halo to the bulge in the initial conditions. Hence,
a deeper potential well was formed. This induced a higher gas mass concentration in the centre
of the galaxy, which increased the star formation. \\

- After 2 Gyr of evolution in isolation, the spiral structure of the ISM developed into concentric rings in the simulation.
When ram pressure was acting on a galaxy, however, the ISM was compressed, which by
itself caused an enhancement of the star formation rate. Moreover, the ISM was stripped
and the galaxy can lose most of its gas up to 90\% if there was high ram pressure. In
this case of high ram pressure acting, the process was basically completed within 200 Myr.
Owing to the steeper potential well of the
same galaxy containing a bulge, less gas was stripped as stated by the Gunn \& Gott
criterion \citep{gunn_gott72}. As confirmed by previous investigations, the inclination angle does not have a big
influence on the amount of stripped gas. Only if the inclination was nearly edge-on
the fraction of gas in the wake of the galaxy was much lower. 
A difference in the amount of stripped gas for
distinct inclination angles can be found when lower ram pressure is acting on the galaxies. \\

- Star formation was enhanced when ram pressure was acting on a galaxy.
In the considered ram-pressure scenarios, the star formation rate was up to four
times higher than for the same galaxy evolving in vacuum. The higher the ram pressure,
the more gas was stripped and therefore fewer stars were formed in the disc. On the other
hand, many more dense star forming gas knots were present in the wake, and the higher
the external pressure on these knots, the more stars were formed. \\

- When the inclination angle of a galaxy was modified, the star formation rate did not change
significantly when low ram pressure acted on a galaxy. Still, the highest SFR could be found 
when the galaxy was flying face-on through the ICM. In the high-density ICM, increasing 
the inclination angle resulted in a slight decrease 
of the star formation rate. Furthermore, in the edge-on case the star formation
rate was lower at the beginning, but higher at the end of the simulation. \\

- The mass fraction of newly formed stars in the wake compared to the total mass was much higher 
for stronger ram pressure. Again, because less gas was stripped if the bulge mass was higher,
fewer stars were formed in the wake when a big bulge was present. In agreement with
previous results, less mass of newly formed stars was present in the wake
after 500 Myr of evolution if the bulge size of a galaxy was increased, and obviously also
the mass in the wake was higher when the ram pressure was increased. \\

- Different disc gas mass fractions do have a direct influence on the star formation 
rate, which decreases with a lower disc gas mass fraction. Interestingly, the enhancement due 
to ram-pressure was less pronounced for larger gas mass fractions. Apparently,
because the gas is less dense when the gas mass fraction is low, ram-pressure can compress
the gas to bring it into a star-forming regime. On the other hand, when the gas mass fraction and 
hence the density was already high, the compression effect was smaller. \\

- Our investigations show that for high ram pressure up to 25\% of the newly
formed stars were formed in the wake of the galaxy, whereas with low ram pressure, only a
marginal amount of around 1\% of all newly formed stars were present in the wake. When
the ICM density was high, gas knots were formed from the stripped material in the wake.
These high dense structures contain a cool core and hence many new stars were formed in
these self-gravitating gas knots. With evolving time the gas knots are losing mass due to star formation processes. 
Although the gas knots were heated by stellar feedback, and 
gas flows out via stellar winds, the gas knots remained intact. Obviously, the cooling was very
efficient. Additionally, self-gravity and the external pressure from the ICM kept the gas knots dense. Hence, the mass 
loss was almost completely due to star formation. The stars in these knots were not subject to ram pressure
and were attracted by the galaxy itself. Because the stars are collisionless, they fall
through the disc and back again, oscillating. Therefore, stars were present
also in front of the disc. \\

- The presence of these gas knots corresponds well to observations of the 'fireballs' in the wake 
of RB199 in the Coma cluster by \citet{yoshida08} and the dense blue gas knots in the wake of IC 3418 in the 
Virgo cluster \citep{martin05}. In both cases, the most plausible mechanism to form these gas knots is 
ram-pressure stripping, also confirmed by the simulations we performed here. \\

\begin{acknowledgements}

The authors thank the anonymous referee, who helped to improve the quality of the paper significantly.
The authors acknowledge the UniInfrastrukturprogramm des BMWF Forschungsprojekt Konsortium Hochleistungsrechnen,
the Austrian Science Foundation FWF through grant P19300-N16 and the doctoral school - Computational Interdisciplinary Modelling
FWF DK-plus (W1227). Furthermore, the authors thank Volker Springel for providing the simulation code GADGET-2 and the initial
conditions generator. We furthermore acknowledge profitable discussion with A. B\"ohm and the colleagues at the institute.

\end{acknowledgements}

\bibliographystyle{aa}
\bibliography{literature.bib}

\begin{thebibliography}{56}
\expandafter\ifx\csname natexlab\endcsname\relax\def\natexlab#1{#1}\fi

\bibitem[{{Abadi} {et~al.}(1999){Abadi}, {Moore}, \& {Bower}}]{abadi99}
{Abadi}, M.~G., {Moore}, B., \& {Bower}, R.~G. 1999, \mnras, 308, 947

\bibitem[{{Barkhouse} {et~al.}(2009){Barkhouse}, {Yee}, \&
  {L{\'o}pez-Cruz}}]{bark09}
{Barkhouse}, W.~A., {Yee}, H.~K.~C., \& {L{\'o}pez-Cruz}, O. 2009, \apj, 703,
  2024

\bibitem[{{Barnes} \& {Hut}(1986)}]{bh86}
{Barnes}, J. \& {Hut}, P. 1986, \nat, 324, 446

\bibitem[{{Bekki} \& {Couch}(2003)}]{bekki03}
{Bekki}, K. \& {Couch}, W.~J. 2003, \apjl, 596, L13

\bibitem[{{Bekki} \& {Couch}(2011)}]{bekki11}
{Bekki}, K. \& {Couch}, W.~J. 2011, \mnras, 415, 1783

\bibitem[{{B\"osch} {et~al.}(2012){B\"osch}, {B\"ohm}, {Wolf},
  {Arag{\'o}n-Salamanca}, {Barden}, {Gray}, {Ziegler}, {Schindler}, \&
  {Balogh}}]{boesch12}
{B\"osch}, B., {B\"ohm}, A., {Wolf}, C., {et~al.} 2012, {\aap $\;$ submitted}

\bibitem[{{Boselli} {et~al.}(2009){Boselli}, {Boissier}, {Cortese}, \&
  {Gavazzi}}]{boselli09}
{Boselli}, A., {Boissier}, S., {Cortese}, L., \& {Gavazzi}, G. 2009,
  Astronomische Nachrichten, 330, 904

\bibitem[{{Bushouse}(1987)}]{bushouse87}
{Bushouse}, H.~A. 1987, \apj, 320, 49

\bibitem[{{Butcher} \& {Oemler}(1978)}]{bo78}
{Butcher}, H. \& {Oemler}, Jr., A. 1978, \apj, 219, 18

\bibitem[{{Crowl} \& {Kenney}(2006)}]{crowl06}
{Crowl}, H.~H. \& {Kenney}, J.~D.~P. 2006, \apjl, 649, L75

\bibitem[{{Crowl} {et~al.}(2005){Crowl}, {Kenney}, {van Gorkom}, \&
  {Vollmer}}]{crowl05}
{Crowl}, H.~H., {Kenney}, J.~D.~P., {van Gorkom}, J.~H., \& {Vollmer}, B. 2005,
  \aj, 130, 65

\bibitem[{{Dressler} {et~al.}(1997){Dressler}, {Oemler}, {Couch}, {Smail},
  {Ellis}, {Barger}, {Butcher}, {Poggianti}, \& {Sharples}}]{dressler97}
{Dressler}, A., {Oemler}, Jr., A., {Couch}, W.~J., {et~al.} 1997, \apj, 490,
  577

\bibitem[{{Durret} {et~al.}(2011){Durret}, {Lagan{\'a}}, \&
  {Haider}}]{durret11}
{Durret}, F., {Lagan{\'a}}, T.~F., \& {Haider}, M. 2011, \aap, 529, A38+

\bibitem[{{Gavazzi} {et~al.}(2006){Gavazzi}, {Boselli}, {Cortese}, {Arosio},
  {Gallazzi}, {Pedotti}, \& {Carrasco}}]{gavazzi06}
{Gavazzi}, G., {Boselli}, A., {Cortese}, L., {et~al.} 2006, \aap, 446, 839

\bibitem[{{Gavazzi} {et~al.}(2001){Gavazzi}, {Marcelin}, {Boselli}, {Amram},
  {V{\'{\i}}lchez}, {Iglesias-Paramo}, \& {Tarenghi}}]{gavazzi01}
{Gavazzi}, G., {Marcelin}, M., {Boselli}, A., {et~al.} 2001, \aap, 377, 745

\bibitem[{{Gingold} \& {Monaghan}(1977)}]{gingold77}
{Gingold}, R.~A. \& {Monaghan}, J.~J. 1977, \mnras, 181, 375

\bibitem[{{Gunn} \& {Gott}(1972)}]{gunn_gott72}
{Gunn}, J.~E. \& {Gott}, III, J.~R. 1972, \apj, 176, 1

\bibitem[{{He{\ss}} \& {Springel}(2010)}]{hess10}
{He{\ss}}, S. \& {Springel}, V. 2010, \mnras, 406, 2289

\bibitem[{{Hester} {et~al.}(2010){Hester}, {Seibert}, {Neill}, {Wyder}, {Gil de
  Paz}, {Madore}, {Martin}, {Schiminovich}, \& {Rich}}]{hester10}
{Hester}, J.~A., {Seibert}, M., {Neill}, J.~D., {et~al.} 2010, \apjl, 716, L14

\bibitem[{{J{\'a}chym} {et~al.}(2009){J{\'a}chym}, {K{\"o}ppen}, {Palou{\v s}},
  \& {Combes}}]{jachym09}
{J{\'a}chym}, P., {K{\"o}ppen}, J., {Palou{\v s}}, J., \& {Combes}, F. 2009,
  \aap, 500, 693

\bibitem[{{J{\'a}chym} {et~al.}(2007){J{\'a}chym}, {Palou{\v s}}, {K{\"o}ppen},
  \& {Combes}}]{jachym07}
{J{\'a}chym}, P., {Palou{\v s}}, J., {K{\"o}ppen}, J., \& {Combes}, F. 2007,
  \aap, 472, 5

\bibitem[{{Kapferer} {et~al.}(2006){Kapferer}, {Kronberger}, {Schindler},
  {B{\"o}hm}, \& {Ziegler}}]{kapferer06}
{Kapferer}, W., {Kronberger}, T., {Schindler}, S., {B{\"o}hm}, A., \&
  {Ziegler}, B.~L. 2006, \aap, 446, 847

\bibitem[{{Kapferer} {et~al.}(2010){Kapferer}, {Schindler}, {Knollmann}, \&
  {van Kampen}}]{kapf10}
{Kapferer}, W., {Schindler}, S., {Knollmann}, S.~R., \& {van Kampen}, E. 2010,
  \aap, 516, A41+

\bibitem[{{Kapferer} {et~al.}(2009){Kapferer}, {Sluka}, {Schindler}, {Ferrari},
  \& {Ziegler}}]{kapf09}
{Kapferer}, W., {Sluka}, C., {Schindler}, S., {Ferrari}, C., \& {Ziegler}, B.
  2009, \aap, 499, 87

\bibitem[{{Katz} {et~al.}(1996){Katz}, {Weinberg}, \& {Hernquist}}]{katz96}
{Katz}, N., {Weinberg}, D.~H., \& {Hernquist}, L. 1996, \apjs, 105, 19

\bibitem[{{Kenney} {et~al.}(2004){Kenney}, {van Gorkom}, \&
  {Vollmer}}]{kenney04}
{Kenney}, J.~D.~P., {van Gorkom}, J.~H., \& {Vollmer}, B. 2004, \aj, 127, 3361

\bibitem[{{Kewley} {et~al.}(2006){Kewley}, {Geller}, \& {Barton}}]{kewley06}
{Kewley}, L.~J., {Geller}, M.~J., \& {Barton}, E.~J. 2006, \aj, 131, 2004

\bibitem[{{Kronberger} {et~al.}(2008){Kronberger}, {Kapferer}, {Ferrari},
  {Unterguggenberger}, \& {Schindler}}]{kronberger08}
{Kronberger}, T., {Kapferer}, W., {Ferrari}, C., {Unterguggenberger}, S., \&
  {Schindler}, S. 2008, \aap, 481, 337

\bibitem[{{Kronberger} {et~al.}(2006){Kronberger}, {Kapferer}, {Schindler},
  {B{\"o}hm}, {Kutdemir}, \& {Ziegler}}]{kronberger06}
{Kronberger}, T., {Kapferer}, W., {Schindler}, S., {et~al.} 2006, \aap, 458, 69

\bibitem[{{Kronberger} {et~al.}(2007){Kronberger}, {Kapferer}, {Schindler}, \&
  {Ziegler}}]{kronberger07}
{Kronberger}, T., {Kapferer}, W., {Schindler}, S., \& {Ziegler}, B.~L. 2007,
  \aap, 473, 761

\bibitem[{{Larson} {et~al.}(1980){Larson}, {Tinsley}, \& {Caldwell}}]{larson80}
{Larson}, R.~B., {Tinsley}, B.~M., \& {Caldwell}, C.~N. 1980, \apj, 237, 692

\bibitem[{{Lovisari} {et~al.}(2009){Lovisari}, {Kapferer}, {Schindler}, \&
  {Ferrari}}]{lovisari09}
{Lovisari}, L., {Kapferer}, W., {Schindler}, S., \& {Ferrari}, C. 2009, \aap,
  508, 191

\bibitem[{{Lucy}(1977)}]{lucy77}
{Lucy}, L.~B. 1977, \aj, 82, 1013

\bibitem[{{Martin}(1999)}]{martin99}
{Martin}, C.~L. 1999, \apj, 513, 156

\bibitem[{{Martin} {et~al.}(2005){Martin}, {Fanson}, {Schiminovich},
  {Morrissey}, {Friedman}, {Barlow}, {Conrow}, {Grange}, {Jelinsky},
  {Milliard}, {Siegmund}, {Bianchi}, {Byun}, {Donas}, {Forster}, {Heckman},
  {Lee}, {Madore}, {Malina}, {Neff}, {Rich}, {Small}, {Surber}, {Szalay},
  {Welsh}, \& {Wyder}}]{martin05}
{Martin}, D.~C., {Fanson}, J., {Schiminovich}, D., {et~al.} 2005, \apjl, 619,
  L1

\bibitem[{{Mo} {et~al.}(1998){Mo}, {Mao}, \& {White}}]{mmw98}
{Mo}, H.~J., {Mao}, S., \& {White}, S.~D.~M. 1998, \mnras, 295, 319

\bibitem[{{Moore} {et~al.}(1998){Moore}, {Lake}, \& {Katz}}]{moore98}
{Moore}, B., {Lake}, G., \& {Katz}, N. 1998, \apj, 495, 139

\bibitem[{{Poggianti} {et~al.}(2006){Poggianti}, {von der Linden}, {De Lucia},
  {Desai}, {Simard}, {Halliday}, {Arag{\'o}n-Salamanca}, {Bower}, {Varela},
  {Best}, {Clowe}, {Dalcanton}, {Jablonka}, {Milvang-Jensen}, {Pello},
  {Rudnick}, {Saglia}, {White}, \& {Zaritsky}}]{poggianti06}
{Poggianti}, B.~M., {von der Linden}, A., {De Lucia}, G., {et~al.} 2006, \apj,
  642, 188

\bibitem[{{Read} {et~al.}(2010){Read}, {Hayfield}, \& {Agertz}}]{read10}
{Read}, J.~I., {Hayfield}, T., \& {Agertz}, O. 2010, \mnras, 405, 1513

\bibitem[{{Roediger} \& {Br{\"u}ggen}(2006)}]{roediger06}
{Roediger}, E. \& {Br{\"u}ggen}, M. 2006, \mnras, 369, 567

\bibitem[{{Roediger} \& {Br{\"u}ggen}(2007)}]{roediger07}
{Roediger}, E. \& {Br{\"u}ggen}, M. 2007, \mnras, 380, 1399

\bibitem[{{Roediger} \& {Hensler}(2005)}]{roediger05}
{Roediger}, E. \& {Hensler}, G. 2005, \aap, 433, 875

\bibitem[{{Rubin} {et~al.}(1999){Rubin}, {Waterman}, \& {Kenney}}]{rubin99}
{Rubin}, V.~C., {Waterman}, A.~H., \& {Kenney}, J.~D.~P. 1999, \aj, 118, 236

\bibitem[{{Schindler} \& {Diaferio}(2008)}]{schindler08}
{Schindler}, S. \& {Diaferio}, A. 2008, Space Science Reviews, 134, 363

\bibitem[{{Scott} {et~al.}(2010){Scott}, {Bravo-Alfaro}, {Brinks}, {Caretta},
  {Cortese}, {Boselli}, {Hardcastle}, {Croston}, \& {Plauchu}}]{scott10}
{Scott}, T.~C., {Bravo-Alfaro}, H., {Brinks}, E., {et~al.} 2010, \mnras, 403,
  1175

\bibitem[{{Solomon} \& {Sage}(1988)}]{sosa88}
{Solomon}, P.~M. \& {Sage}, L.~J. 1988, \apj, 334, 613

\bibitem[{{Springel}(2005)}]{springel05}
{Springel}, V. 2005, \mnras, 364, 1105

\bibitem[{{Springel}(2010)}]{springel10}
{Springel}, V. 2010, \mnras, 401, 791

\bibitem[{{Springel} \& {Hernquist}(2003)}]{springel03}
{Springel}, V. \& {Hernquist}, L. 2003, \mnras, 339, 289

\bibitem[{{Sulentic}(1976)}]{sulentic76}
{Sulentic}, J.~W. 1976, \apjs, 32, 171

\bibitem[{{Tran} {et~al.}(2005){Tran}, {van Dokkum}, {Illingworth}, {Kelson},
  {Gonzalez}, \& {Franx}}]{tran05}
{Tran}, K.-V.~H., {van Dokkum}, P., {Illingworth}, G.~D., {et~al.} 2005, \apj,
  619, 134

\bibitem[{{Vollmer} {et~al.}(2001){Vollmer}, {Cayatte}, {Balkowski}, \&
  {Duschl}}]{vollmer01}
{Vollmer}, B., {Cayatte}, V., {Balkowski}, C., \& {Duschl}, W.~J. 2001, \apj,
  561, 708

\bibitem[{{Vollmer} {et~al.}(2012){Vollmer}, {Soida}, {Braine}, {Abramson},
  {Beck}, {Chung}, {Crowl}, {Kenney}, \& {van Gorkom}}]{vollmer12}
{Vollmer}, B., {Soida}, M., {Braine}, J., {et~al.} 2012, \aap, 537, A143

\bibitem[{{Wolf} {et~al.}(2009){Wolf}, {Arag{\'o}n-Salamanca}, {Balogh},
  {Barden}, {Bell}, {Gray}, {Peng}, {Bacon}, {Barazza}, {B{\"o}hm}, {Caldwell},
  {Gallazzi}, {H{\"a}u{\ss}ler}, {Heymans}, {Jahnke}, {Jogee}, {van Kampen},
  {Lane}, {McIntosh}, {Meisenheimer}, {Papovich}, {S{\'a}nchez}, {Taylor},
  {Wisotzki}, \& {Zheng}}]{wolf09}
{Wolf}, C., {Arag{\'o}n-Salamanca}, A., {Balogh}, M., {et~al.} 2009, \mnras,
  393, 1302

\bibitem[{{Yoshida} {et~al.}(2004){Yoshida}, {Ohyama}, {Iye}, {Aoki},
  {Kashikawa}, {Sasaki}, {Shimasaku}, {Yagi}, {Okamura}, {Doi}, {Furusawa},
  {Hamabe}, {Kimura}, {Komiyama}, {Miyazaki}, {Miyazaki}, {Nakata}, {Ouchi},
  {Sekiguchi}, \& {Yasuda}}]{yoshida04}
{Yoshida}, M., {Ohyama}, Y., {Iye}, M., {et~al.} 2004, \aj, 127, 90

\bibitem[{{Yoshida} {et~al.}(2008){Yoshida}, {Yagi}, {Komiyama}, {Furusawa},
  {Kashikawa}, {Koyama}, {Yamanoi}, {Hattori}, \& {Okamura}}]{yoshida08}
{Yoshida}, M., {Yagi}, M., {Komiyama}, Y., {et~al.} 2008, \apj, 688, 918

\end{thebibliography}

\end{document}